\documentclass[12pt]{article}

\usepackage{microtype}
\usepackage{graphicx}
\usepackage{subfigure}
\usepackage{booktabs} 
\usepackage[margin=25mm]{geometry}
\usepackage{hyperref}

\providecommand{\keywords}[1]
{
  \small	
  \textbf{\textit{Keywords---}} #1
}

\usepackage{enumitem}
\usepackage{amsmath}
\usepackage{amssymb}
\usepackage{amsthm}
\usepackage{bm}
\usepackage{tablefootnote}
\usepackage{tabularx}
\usepackage{threeparttable}
\usepackage{natbib}
\usepackage[titletoc]{appendix}
\newtheorem{Theorem}{Theorem}

\newtheorem{assumption}{Assumption}

\begin{document}

\title{Causal Effect Estimation and Optimal Dose Suggestions in Mobile Health}




\author{Liangyu Zhu$^1$\footnote{Email: lzhu12@ncsu.edu},
        Wenbin Lu$^1$, 
        Rui Song$^1$\\
        $^1$Department of Statistics, North Carolina State University}
\date{}

\maketitle




\begin{abstract}
In this article, we propose novel structural nested models to estimate causal effects of continuous treatments based on mobile health data. To find the treatment regime that optimizes the expected short-term outcomes for patients, we define a weighted lag-$K$ advantage as the value function. The optimal treatment regime is then defined to be the one that maximizes the value function. Our method imposes minimal assumptions on the data generating process. Statistical inference is provided for the estimated parameters. Simulation studies and an application to the Ohio type 1 diabetes dataset show that our method could provide meaningful insights for dose suggestions with mobile health data. 
\end{abstract}

\vskip 0.3in

\section{Introduction}

There is a rapid-increasing interest in healthcare interventions using mobile apps. Mobile technologies allow physical conditions of patients to be collected in real time, measured by sensors or self-reported by patients. Studies have shown that mobile health interventions could be beneficial for the healthcare delivery process by improving disease management, enhancing communication with the healthcare provider and providing more precise and individualized medication \citep{free2013effectiveness}. However, analyzing mobile health data can be challenging because they typically have a large number of time points,  time-varying treatments, and non-definite time horizon \citep{luckett2019estimating}.

One focus in analyzing mobile health data is to evaluate causal effects of mobile health interventions. Generalized estimating equations (GEE) are commonly used for studying dependence of an outcome variable on a set of covariates observed over time \citep{liang1986longitudinal,zhao1990correlated,liang1992multivariate,schafer2006marginal}. GEE enhance the efficiency of the generalized linear models by including into the estimation equations the correlations among repeated observations of a subject over time. Such approaches typically require a full working correlation model and will be computationally expensive as the time points get larger. Application of GEE in mobile health data has been limited to time-invariant treatments \citep{evans2012pilot,carra2016impact}. \citet{liao2015micro} proposed the micro-randomized trial design for estimating the causal effect of just-in-time treatments under the mobile health setting \citep{klasnja2015microrandomized,liao2016sample,dempsey2015randomised}. \citet{liao2016sample} and \citet{boruvka2018assessing} defined the proximal and lagged treatment effects for time-varying treatments with data from micro-randomized trials.  A centered and weighted estimation method based on inverse probability of treatment-estimators \citep{robins2000marginal,murphy2001marginal} is then proposed for estimating these causal effects. 

Providing personalized treatment suggestions based on mobile health data is also of great interest. Dynamic treatment regimes (DTR) have been proposed for providing sequential treatment suggestions based on longitudinal data from randomized trials or observational data \citep{murphy2003optimal,moodie2007demystifying,kosorok2015adaptive,chakraborty2013statistical}. A dynamic treatment regime is a set of decision rules that decide treatments to be assigned to patients according to their time-varying measurements during the ongoing treatment process. An optimal DTR is the one that yields the most favorable expected mean outcome over a fixed period of time. Optimal dynamic treatment regimes are typically estimated by backward induction based on parametric models for the expected outcome (Q-learning) \citep{watkins1992q,sutton1998introduction,murphy2005experimental,schulte2014q}. Robustness of these methods can be further enhanced by using semi-parametric models \citep{murphy2003optimal,robins2004optimal,moodie2007demystifying,tang2012developing,schulte2014q} or non-parametric models \citep{zhao2009reinforcement}. \citet{zhao2015new} avoid the risk of model misspecification by directly maximizing a nonparametric estimation of the cumulative reward among a predefined class of treatment regimes. 

However, mobile health data usually have infinite time horizons. Sequential decision making process in infinite horizon can be modeled as a Markov decision process \citep{puterman2014markov}. \citet{ertefaie2018constructing} defined the optimal DTR in the infinite horizon as the one which maximizes the expected cumulative discounted reward (the beneficial outcome). The optimal DTR is estimated first by positing a parametric model for the maximum expected cumulative discounted reward. Least square estimation equations are then constructed based on the Bellman equation \citep{sutton1998introduction}. The optimization is achieved through greedy gradient Q-learning \citep{maei2010toward}. \citet{luckett2019estimating} proposed the V-learning method for finding the optimal DTR. They first posit a model for the expected cumulative discounted reward of a specific treatment regime. Then they search for the treatment regime which maximizes the estimated cumulative discounted reward function within a prespecified class of treatment regimes. However, both of these two methods are limited to discrete treatments.
 
There is increasing attention in how mobile interventions can help managing diseases by monitoring physical conditions against high-risk events and providing frequent treatment adjustments \citep{maahs2012outpatient,levine2001predictors}. Our research is motivated by the use of mobile health applications for diseases like diabetes and hypertension, where the main focus is to monitor adverse events in the near future \citep{haller2004predictors,heron2010ecological}. For example, for diabetes patients, the main interest of using rapid-reacting insulin is to maintain a safe blood glucose level within 2 hours after a meal. Existing methodologies in reinforcement learning mainly aims at maximizing a discounted cumulative reward, which might not be the optimal criteria for treatment suggestions in this scenario. Furthermore, the treatments in this case are continuous and thus have infinite number of possible values. Studies on estimating causal effects and providing treatment suggestions under this setting are still absent. 

In this article, we aim to find the treatment regime which optimizes the outcomes (or minimizes the risk of adverse events) within a time period of near future. We first extend \citet{boruvka2018assessing}'s definition of lagged treatment effect to continuous treatments and propose novel structural nested models for estimating causal effects of continuous treatments based on mobile health data. We then define a weighted advantage function. The optimal treatment regime at a specific time point is defined to be the one which optimizes the weighted advantage function. The rest of the article is structured as follows. In section 2, we formalize the problem in a statistical framework and present the proposed methodology for finding the optimal DTR. In section 3, we discuss the theoretical results of the proposed estimators. Simulations are conducted and the corresponding results are presented in section 4. In section 5, we apply the proposed method to the Ohio type 1 diabetes dataset. Discussions and conclusions are given in section 6.

\section{Method}
\subsection{Notation}
We assume that for each individual, the measurements are taken at time points with fixed time intervals, $t=1,...,T$. Let $A_t\in\mathcal {A}$ denotes the treatment at decision time $t$, where $\mathcal A$ is a continuous interval of possible values of doses.  $X_t\in\mathbb{R}^p$ are covariates measured at time $t$. $Y_{t}\in\mathbb{R}$ denotes the outcome measured at time $t$ following the decision $A_{t-1}$, $t>1$.  Without loss of generality, we assume that higher values of $Y_{t}$ denote better outcomes. We assume that $X_t$ and $Y_t$ are observed simultaneously and $A_t$ is a decision made after observing $X_t$ and $Y_t$. Thus, the observed data for one subject are $\{(X_1,A_1)$,$(Y_2,X_2,A_2)$, $\dots$, $(Y_T, X_T,A_T)$, $(Y_{T+1},X_{T+1})\}$. In this article, we use capitalized letters to denote random variables and lowercase letters to denote realized values. Let the overbar denotes the history of a random variable. For example, $\bar X_{t}=(X_1,...,X_t)$. All information accrued up to time $t$ can be represented by $H_t=(\bar X_t , \bar Y_t,\bar A_{t-1})$. In the considered type-1 diabetes study, $A_t$ is the rapid-reacting insulin dose taken at time $t$. $Y_{t}$ measures the stability of the blood glucose between time $t-1$ and $t$, and $X_t$ includes the food intake, exercise and blood glucose levels.

To define the treatment effects, we adopt the potential outcome framework by \citet{rubin1974estimating}.
$X_t(\bar a_{t-1})$ and $Y_t(\bar a_{t-1})$ are the potential measurements of covariates and potential outcomes at time $t$ had the sequence of treatments $\bar a_{t-1}$ been allocated to the patient, $\bar a_{t-1}\in \mathcal A^{t-1}$. $A_t(\bar a_{t-1})$ is defined as the potential treatment at $t$ had the sequence of $\bar a_{t-1}$ be allocated. This notation implicitly assumes that the potential outcomes are not influenced by future treatments and the outcome of one subject is not affected by the treatments received by other subjects. The latter is also known as the stable unit treatment value assumption (SUTVA; see \citet{rubin1974estimating}). 
For simplicity, we denote $A_2(A_1)$ by $A_2$, $A_t(\bar A_{t-1})$ by $A_t$. Then $H_t(\bar A_{t-1})=\{X_1$, $A_1$, $Y_2(A_1)$, $X_2(A_1)$, $A_2(A_1)$, $\dots$, $Y_t(\bar A_{t-1})$, $X_t(\bar A_{t-1})\}$. 

A dynamic treatment regime $\pi=(\pi_1,...,\pi_T)$ is a set of rules that outputs a distribution of treatment options at each time point based on past history $\pi_t=\{ p_{\pi,t}(a|h_t)$, $a \in \mathcal A \}$;  $p_{\pi,t}$ here denotes the conditional density of choosing treatment $a$ given history $h_t$ at time $t$. Let $\mathcal H_t$ be the space of all possible histories. A treatment regime is deterministic if $p_{\pi,t}(a|h_t)=\delta(a=g_t(h_t))$, for some $g_t:\mathcal H_t \to \mathcal A$, where $\delta(\cdot)$ is the Dirac delta function. Then for simplicity of notation, we write $\pi_t$ as $\pi_t=g_t(h_t)$. 
\subsection{Lag k Treatment Effect}

Define the conditional lag $k$ $(k\ge 1)$ treatment effect of treatment $a$ with respect to a reference treatment $a_0$ at time $t$ as:
\begin{align}
    &\tau_{t,k}\big(a,a_0,H_t(\bar A_{t-1})\big)=\nonumber\\
    &E\big\{Y_{t+k}(\bar A_{t-1}, a , A_{t+1}^{a_t=a},\dots, A_{t+k-1}^{a_t=a})-
    &Y_{t+k}(\bar A_{t-1}, a_0, A_{t+1}^{a_t=a_0}, \dots, A_{t+k-1}^{a_t=a_0})|H_{t}(\bar A_{t-1})\big\}. \label{lagk-effect}
\end{align}
 $A_{t+1}^{a_t=a}$ denotes the potential treatment $A_{t+1}(\bar A_{t-1},A_t=a)$, $A_{t+l}^{a_t=a}$ denotes $A_{t+l}(\bar A_{t-1},$ $A_t=a,$ $ A_{t+1}^{a_t=a},$ $...,$ $A_{t+l-1}^{a_t=a})$, for $l=2,\dots,k-1$. The expectation in Equation (\ref{lagk-effect}) is taken over all the possible future treatments from time $t$ to $t+k-1$. Notice that the treatment effect defined in (\ref{lagk-effect}) measures the effect of a one-time change in the decision strategy. The causal effect measuring a single-time decision change has been extensively used in various models for intensively collected longitudinal data \citep{schafer2006marginal, schwartz2007analysis, bolger2013intensive}.  \citet{boruvka2018assessing} also used a similar definition for estimating the effect of mobile application notifications.

To use the observed data to estimate the lag $k$ treatment effect, we make the following assumptions \citep{robins2004optimal}:
\begin{enumerate}[wide, labelindent=0pt]
    \item Consistency: The potential outcomes had the treatments given to the patient equal to the observed treatment history are equal to the observed data. More specifically, for $\bar a_{t-1}=\bar A_{t-1}$, $ Y_t(\bar a_{t-1})= Y_t$, $ X_t(\bar a_{t-1})= X_t$ and $ A_t(\bar a_{t-1})= A_t$ for $2\le t\le T$
    , where the left sides of the equations are the potential outcomes and right sides are the observed variables
    ; At time $T+1$, $\bar Y_{T+1}(\bar a_{T})=\bar Y_{T+1}$, and $\bar X_{T+1}(\bar a_{T})=\bar X_{T+1}$ for $\bar a_{t-1}=\bar A_{t-1}$. 
    
    \item Positivity: All treatments $a\in \mathcal A$ can possibly be observed given $h_t$ for any $h_t\in \mathcal H_t$. More specifically, for $a\in\mathcal A$ and $h_t\in \mathcal H_t$, $p_{\pi,t}(a|h_t)>0$, where $p_{\pi,t}(a|h_t)$ denotes the conditional density for the treatment $A_t$ given the history $H_t = h_t$. In randomized trials, this assumption can be ensured by design. In observational studies, the set of possible treatments may differ by treatment history. The expected outcomes of a treatment regime $\pi$ cannot be estimated from the observed data when it has a non-zero probability of suggesting treatments where $p_{\pi,t}(a|h_t)>0$ and $p(A_t=a|H_t=h_t)=0$. Such treatment regimes are regarded as infeasible treatment regimes \citep{robins2004optimal}. We could then limit our attention to feasible treatment regimes by adding constraints to the suggested treatments. In practice, this assumption is hard to examine when $\mathcal A$ is unknown. In such scenarios, causal inference for treatments that are uncommon in the dataset should be made with caution.   
    
    \item Sequential ignorability: The potentials outcomes $\{Y_{t+1}(\bar a_t),X_{t+1}(\bar a_t), A_{t+1}(\bar a_t), ...,Y_{T+1}(\bar a_T)\}$ are independent of $A_t$ conditional on $H_t$, for $t\le T$. This assumption is naturally satisfied in a sequentially randomized study, where treatments are randomized for each time point. In an observational study, this assumption cannot be verified and is often assumed. 
\end{enumerate}

Under these three assumptions, we can estimate the conditional lag $k$ treatment effect with the observed data for any $a\in \mathcal A$ (see appendix for the proof):
\begin{align}\label{potential-observed}
    &E\big\{Y_{t+k}(\bar A_{t-1} , a , A_{t+1}^{a_t=a} , \dots , A_{t+k-1}^{a_t=a} )-
    Y_{t+k}(\bar A_{t-1}, a_0, A_{t+1}^{a_t=a_0}, \dots, A_{t+k-1}^{a_t=a_0})|H_{t}(\bar A_{t-1})\big\} \nonumber\\
    &=E(Y_{t+k}|A_t=a,H_t)-E(Y_{t+k}|A_t=a_0,H_t).
\end{align}
\subsection{Lag K Weighted Advantage}
Furthermore, we define the lag K weighted advantage to be:
$$
    \tilde\tau_{t,K}(a,a_0,S_t(\bar A_{t-1}))=\sum_{k=1}^K w_k\tau_{t,k}(a,a_0,H_t(\bar A_{t-1})),
$$
where $K$ is the largest lag of interest and $w_1,\dots w_K$ are predefined non-negative weights and $w_1+\dots +w_K=1$. For example, if we have hourly data of diabetes patients and we want to minimize the amount of time the blood sugar being outside 80-140 mg/dL within four hours after the dose injection, we could define $Y_t$ as the percentage of time the blood sugar being outside the optimal range at the $t$-th hour. Take $K=4$ and $w_1=w_2=w_3=w_4=0.25$. An optimal dose suggestion at time $t$ would be the one which maximizes the lag $K$ weighted advantage at time $t$. Therefore, we define the optimal treatment regime at time $t$ to be:
$
    \pi_t^{opt}=\arg\max_{\pi_t} \tilde{\tau}_{t,K}\big \{a=\pi_t(H_t(\bar A_{t-1})),a_0, H_t(\bar A_{t-1})\big \}
$. 
Notice that the choice of $a_0$ does not affect the optimal treatment regime. In this article, we take $a_0=0$
. For simplicity of notation, we write $\tau_{t,k}(a,0,H_t(\bar A_{t-1}))$ as $\tau_{t,k}(a,H_t(\bar A_{t-1}))$ in the rest of this article.

\subsection{Estimation Method}
We first use a nonparametric version of the structural nested models to estimate the lag $k$ treatment effect. 
The following model assumes that the lag $k$ treatment effects for $k=1,\dots,K$ depend on $H_t$ only through $S_t$, where $S_t\in\mathcal S$ are some summary statistics of the past history.
\begin{align}\label{model}
    &\tau_{t,k}(a,H_t)=\tau_k(a,S_{t};\alpha_k,\beta_k) =\alpha_k a^2+  \{\beta_k^T f_k(S_{t}) \}a.
\end{align}
where $f_k$ is a $q_k$ dimensional function of $S_t$. Notice that we assume $S_t$ for $t=1,\dots, T$ to be from the same vector space $\mathcal S$ and the parameters in this model do not vary with $t$.  \citet{boruvka2018assessing} showed that the models for the lagged effects for different $k$ do not constrain one another. 
The motivation for using a quadratic model is that both underdosing and overdosing might lead to unfavorable outcomes in practice. Let $\alpha=(\alpha_1,\dots,\alpha_K)^T, \beta=(\beta_1,\dots,\beta_K)^T$, and $w=(w_1,...,w_K)^T$. Without loss of generality, we assume that $f_1(S_t)=\dots=f_K(S_t)$. (Otherwise, just let $f(S_t)$ be a vector of functions which includes all the functions from $\{f_1(S_t),\dots,f_K(S_t)\}$ and substitute $f_1(S_t),\dots,f_K(S_t)$ with $f(S_t)$.) Then the weighted lag $K$ advantage is:
\begin{align*}
    &\tilde \tau_{K}(a,S_t;\alpha,\beta)=\sum_{k=1}^K w_k \tau_k(a,S_t,\alpha_k,\beta_k)\\
    &=\{w^T\alpha\} a^2+\{w^T\beta\}^T f(S_t) a
    =\tilde \alpha_K a^2 +\tilde \beta_K^T f(S_t) a,
\end{align*}
where $\tilde \alpha_K=w^T\alpha$, $\tilde \beta_K=w^T\beta$. Thus the lag $K$ weighted advantage also follows a quadratic form. Notice that under the model above, $\tilde \tau_{t,K}\big(a,a_0,H_t(\bar A_{t-1})\big)=\tilde \tau_{K}(a,S_t; \alpha,\beta)$ also depends on $H_t$ only through $S_t$. Thus the optimal treatment regime 
$\pi_t^{opt}=\arg\max_{\pi_t} \tilde \tau_{K}(a=\pi_t,S_t;\alpha,\beta)$ also depends on $H_t$ only through $S_t$. When $\tilde \alpha_K<0$, the optimal dose at time $t$ would be a deterministic treatment regime: $\pi_t^{opt}=-\{\tilde \beta_K^T f(S_t)\}/2\tilde\alpha_K$. The parameter $-\tilde\beta_{K,j}/\tilde \alpha_K$ can be interpreted as the difference of the optimal dosage for patients with one unit difference in the $j$-th term of $f(S_t)$ while having all the other covariates the same, $j=1,\dots,q$ where $q$ is the dimension of $f(S_t)$. When $\tilde \alpha_K\ge 0$, the optimal treatment falls on the edge of $\mathcal A$.  

We first present the standard structural nested models for estimating the lag $k$ causal effect. Let $U_{t+k}=Y_{t+k}-\tau_k(A_t, H_t)$. Under the proposed model,
\begin{align*}
    &U_{t+k}(\alpha_k,\beta_k)=Y_{t+k}-\tau_k(A_t,S_{t};\alpha_k, \beta_k)\\&=Y_{t+k}-\alpha_k A_t^2-\beta_k^Tf_k(S_{t}) A_t.
\end{align*}
According to Theorem 3.3 in \citet{robins2004optimal}, under the assumption of sequential randomization and consistency, we can obtain:
\begin{align}\label{unbiasedness0}
    E\big[&\big\{d(A_t,H_t)-E(d(A_t,H_t)|H_t)\big\}\times \big\{ U_{t+k}-E(U_{t+k}|H_t)\big\}\big]=0,
\end{align}
where $d(\cdot,\cdot)$ is an arbitrary function and $t\in\{1,\dots,T+1-k\}$. Assume that the data consist of $n$ independent subjects $\{H^1_{T+1},\dots,H^n_{T+1}\}$
. Then we can estimate $\alpha_k,\beta_k$ with:
\begin{align}\label{snmm-s}
    0=&\mathbb P_n \sum_{t=1}^{T-k+1}\big\{d_{t+k} (A_t,H_{t})-E\big(d_{t+k}(A_t,H_{t})|H_{t}\big)\big\}
    \times \big\{U_{t+k}(\alpha_k, \beta_k)-E\big(U_{t+k}(\alpha_k,\beta_k)|H_{t}\big)\big\},
\end{align}
where
\begin{align}\label{d}
    d_{t+k}(A_t,H_{t})=-\frac{\partial U_{t+k}(\alpha_k,\beta_k)}{\partial(\alpha_k, \beta_k)}=
        \begin{pmatrix}
            A_t^2\\
            A_tf_k(S_{t})
        \end{pmatrix}.
\end{align}
and $\mathbb P_n $ denotes empirical mean of a function, $\mathbb P_n g(A_t,S_t,Y_t)$ $=\sum_{i=1}^n g(A^i_t,S^i_t,Y^i_t)/n$ for any function $g(\cdot)$. Since $d_{t+k}(\cdot)$ depends on $H_t$ only through $S_t$, we write it as $d_{t+k}(A_t,S_t)$. 
To apply the estimation equation, we need to obtain $E\{d_{t+k}(A_t,S_t)|H_t\}$ and $E(Y_{t+k}|H_t)$. The traditional approach is to use regression models to estimate these conditional expectations. However, the complexity of the model increases as $t$ increases, leading to a high risk of model misspecification
. If nonparametric estimators are used, the high dimension of $H_t$ can also induce large variance. Therefore, we revise the estimation equation by first showing the following result (See the appendix for the proof).
\begin{Theorem}\label{strong-unbiasedness}
    Under model assumption (\ref{model}), if the following assumption is satisfied:
\begin{equation}\label{assumption:strong}
    A_t\perp Y_{t+k}(\bar a_{t+k-1})|S_t \quad \text{ for }\quad \bar a_{t+k-1}\in \mathcal A^{t+k-1},
\end{equation}
then 
for an arbitrary function $d(\cdot):\mathcal A\times \mathcal S\to \mathbb R^{q_k+1}$:
\begin{align}\label{unbiasedness}
    E\big[&\big\{d(A_t,S_t)-E(d(A_t,S_t)|S_t)\big\}\times \big\{ U_{t+k}-E(U_{t+k}|S_t)\big\}\big]=0.
\end{align}
\end{Theorem}
Therefore we can estimate $\alpha_k,\beta_k$ with:
\begin{align*}
    0=&\mathbb P_n \sum_{t=1}^{T-k+1}\big\{d_{t+k} (A_t,S_{t})-E\big(d_{t+k}(A_t,S_{t})|S_{t}\big)\big\}\times \big\{U_{t+k}(\alpha_k, \beta_k)-E\big(U_{t+k}(\alpha_k,\beta_k)|S_{t}\big)\big\},
\end{align*}
where $d_{t+k}(A_t,S_t)$ is also taken to be (\ref{d}). The advantage of this estimation equation is that the dimension of $S_t$ does not increase with $t$. Therefore we can use nonparametric estimators for $E(U_{t+k}|S_t)$ and $E(d(A_t,S_t)|S_t)$ without imposing model assumptions on $A_t|S_t$ and $Y_{t+k}|S_t$. The above equation can thus be written as:
\begin{align*}
    &0=\\&\sum_{i=1}^n\sum_{t=1}^{T-k+1}         
        \begin{pmatrix}
            {A_t^i}^2-E({A_t^i}^2|S^i_{t})\\
            \{{A_t^i}-E({A_t^i}|S^i_{t})\}g_k(S^i_{t})
        \end{pmatrix}\Big\{ Y^i_{t+k}-                 
        E( Y^i_{t+k} | S^i_{t} )
        -\begin{pmatrix}
            {A^i_t}^2-E({A^i_t}^2|S^i_{t})\\
            \{A^i_t-E(A^i_t|S^i_{t})\}f_k(S^i_{t})
        \end{pmatrix}^T\begin{pmatrix}
            \alpha_k\\
            \beta_k
        \end{pmatrix}\Big\}.
\end{align*}
Let $B_t(s)=E(A_t^2|S_{t}=s)$, $C_t(s)=E(A_t|S_{t}=s)$, $D_{t,k}(s)=E(Y_{t+k}|S_{t}=s)$. 
We estimate $B_t(s),$ $C_t(s),$ $D_{t,k}(s)$ with kernel estimators:
$
    \hat {B}_{t}(s)=\sum_{i=1}^n {A^i_t}^2 K_{\Lambda}(h-S^i_{t})/\{\sum_{i=1}^n K_{\Lambda}(s-S^i_{t})\}$,
$    \hat C_{t}(s)=\sum_{i=1}^n A_t^i K_{\Lambda}(s-S^i_{t})/ \{\sum_{i=1}^n K_{\Lambda}(s-S^i_{t})\}$,
   $ \hat D_{t,k}(s)=\sum_{i=1}^n Y_{t+k}^i K_{\Lambda}(s-S^i_{t})/ \{\sum_{i=1}^n K_{\Lambda}(s-S^i_{t})\}$,
 where $K(\cdot)$ is a multivariate kernel function and $K_\Lambda(u)=|\Lambda|^{-1/2}K(\Lambda^{-1/2}u)$, $\Lambda$ is a symmetric and positive definite bandwidth matrix. We can then derive the estimated parameters:
 
\begin{small}
\begin{align*}
   &\begin{pmatrix}
            \hat \alpha_k\\
            \hat \beta_k
    \end{pmatrix}
    =\nonumber\\&
    \Big[\sum_{i=1}^n\sum_{t=1}^{T-k+1}  \begin{pmatrix}
            {A_t^i}^2-\hat B_{t}(S_{t}^i)\\
            \{A_t^i-\hat C_{t}(S_{t}^i)\}f_k(S^i_{t})
        \end{pmatrix}^{\otimes 2}\Big]^{-1}
        \Big[\sum_{i=1}^n\sum_{t=1}^{T-k+1}\begin{pmatrix}
            {A_t^i}^2-\hat B_{t}(S_{t}^i)\\
            \{A_t^i-\hat C_{t}(S_{t}^i)\}f_k(S^i_{t})
        \end{pmatrix}\Big\{Y^i_{t+k}-\hat D_{t,k}(S_{t}^i)\Big\}\Big].
\end{align*}
\end{small}

The estimated $\tilde \alpha_K$ and $\tilde \beta_K$ can thus be calculated by $\hat{\tilde \alpha}_K=\sum_{k=1}^Kw_k\hat \alpha_k$, $\hat{\tilde \beta}_K=\sum_{k=1}^K w_k\hat \beta_k$. When $\hat{\tilde \alpha}_K<0$, $\pi_t^{opt}$ can be estimated by $\hat\pi_t^{opt}=-\{\hat{\tilde \beta}_K^T f(S_t)\}/2\hat{\tilde\alpha}_K$. When $\hat{\tilde\alpha}\ge0$, $\hat\pi_t^{opt}$ would be either $0$ or the maximum possible dosage. 
We can also estimate the parameters for the lag $K$ weighted advantage directly by letting $\tilde U_{t+K}(\tilde \alpha_K,\tilde\beta_K)=\sum_{i=1}^Kw_kY_{t+k}-\tilde\alpha_KA_t^2-\{\tilde\beta_K^Tf(S_t)\} A_t$ and estimate $\tilde\alpha_K,\tilde\beta_K$ with:
\small
\begin{align*}
    &0=\sum_{i=1}^n\sum_{t=1}^{T-K+1}\big\{d_{t+K}^i (A_t,S_{t})-E\big(d_{t+K}^i(A_t,S_{t})|S^i_{t}\big)\big\}
    \times\big\{\tilde U_{t+K}^i(\tilde \beta_K)-E\big(\tilde U_{t+K}^i(\tilde \beta_K)|S_{t}\big)\big\},
\end{align*}
\normalsize
where $E\big(\tilde U_{t+K}^i(\tilde \beta_K)|S_t)$ can be estimated similarly with kernel estimation. It is trivial to prove that estimated $\tilde\alpha_K$ and $\tilde \beta_K$ are the same with these two approaches.

Since in model (\ref{model}), the parameters $\alpha_k$ and $\beta_k$ are invariant across time, the estimation equation can thus be summed over the time index $t$. Also notice that the kernel estimation in our method averages over the $n$ observations but not over the time index $t$.  If we include enough information in $S_t$, then it might be possible to assume that the conditional distributions $Y_{t+k}|S_t$ and $A_t|S_t$ are invariant across time. Then we can let:$
    \hat {B}_{t}(s)=\{\sum_{i=1}^n\sum_{t=1}^T {A^i_t}^2 K_{\Lambda}(s-S^i_{t})\}/\{\sum_{i=1}^n\sum_{t=1}^T K_{\Lambda}(s-S^i_{t})\}$, where the sum is taken over $t$ as well (similar for $\hat C_t(s)$ and $\hat D_{t,k}(s)$).
This would be more preferable when we only observe the data of a small number of patients and each patient has a large number of observations over time.

The validity of our estimation equation is mainly based on assumptions (\ref{model}) and (\ref{assumption:strong}). In other words, we assume that the summary statistics of the past history $S_t$ contains all the information which influences the lag $k$ treatment effect and the dependence between $A_t$ and $Y_{t+k}(\bar A_{t-1},a,A_{t+1}^{a_t=a},\dots,A_{t+k-1}^{a_t=a})$ for $k=1,\dots,K$ . In our simulation study, we will also examine the performance of the model when assumption (\ref{assumption:strong}) is not valid. 

\section{Theoretical Results}
In this section, we derive the consistency and asymptotic normality of the estimated parameters. For simplicity of notation, let $B=\{B_1(S_{1}),$ $\dots,$ $B_T(S_{T})\}$, $C=\{C_1(S_{1})$, $\dots,$  $C_T(S_{T})\}$, $D=\{D_1(S_{1}),$ $\dots,$ $D_{T-k+1}(S_{T})\}$, and $\hat B=\{\hat B_1(S_{1}),$ $\dots,$ $\hat B_T(S_{T})\}$, $\hat C=\{\hat C_1(S_{1}),$ $\dots,$ $\hat C_T(S_{T})\}$, $\hat D=\{\hat D_1(S_{1}),$ $\dots,$ $\hat D_{T-k+1}(S_{T})\}$ and $H=H_{T+1}$. Then the solution to the estimating equation can be written as:
\begin{align*}
( \hat\alpha_k, \hat \beta_k^T)^T
    &=\big[\mathbb P_n L_1(H;\hat B,\hat C)\big]^{-1}
    \big[ \mathbb P_n
   L_2(H;\hat B,\hat C,\hat D)
    \big],
\end{align*}
where,
\begin{align*}
     &L_1(H;B,C)= \sum_{t=1}^{T-k+1}   
    \begin{pmatrix}
            {A_t}^2-B_t(S_{t})\\
            \{A_t-C_t(S_{t})\}f_k(S_{t})
    \end{pmatrix}^{\otimes 2},\\
     &L_2(H;B,C,D)=
     \sum_{t=1}^{T-k+1}   
     \begin{pmatrix}
            {A_t}^2-B_t(S_{t})\\
            \{A_t-C_t(S_{t})\}f_k(S_{t})
    \end{pmatrix}
    \Big\{Y_{t+k}-D_t(S_{t})\Big\}.
\end{align*}

Let $\hat \phi_k=(\hat \alpha_k,\hat \beta_k^T)^T$, and $\phi_k^*=(\alpha_k^*,\beta_k^{*T})^T$, where:
\begin{align*}
       \begin{pmatrix}
            \alpha_k^*\\
            \beta_k^{*}
    \end{pmatrix}
    &=\Big\{E\big[L_1(H;B,C)\big]\Big\}^{-1} E\Big[L_2(H;B,C,D)\Big].
\end{align*}
From Equation (\ref{unbiasedness}), it is trivial to obtain that $(\alpha_k^*,\beta_k^{*T})^T$ are the true parameters of the model if the model assumption (\ref{model}) is correct. To derive the asymptotic normality of the estimators, we need the following regularity assumptions:

\begin{assumption}\label{a1}
The marginal density of $S_t$, $p_{S_t}$, is uniformly bounded away from 0 for all $t$: $\inf_{s\in\mathcal S} p_{S_t}(s)>0.$ 
\end{assumption}
\begin{assumption}\label{a2}
As $\Lambda\to 0$, the kernel function satisfies the following equations:\\
$\inf_{s}\{\int_{\mathcal V_s}K(v)dv\}=1-O(\Lambda^{\frac{1}{2}})$; $\sup_{s}\{\int_{\mathcal V_s}vK(v)dv\}=O(1)$;    $\sup_{s}\{\int_{\mathcal V_s}K^2(v)dv\}=O(1)$; $\sup_{s}\{\int_{\mathcal V_s}vK^2(v)dv\}=O(1)$, where $\mathcal V_s=\{v :s-\Lambda^{\frac{1}{2}}v\in \mathcal S\}$ for $s\in\mathcal S$ and $v$ is a vector with the same number of dimensions as $s$.
\end{assumption}
\begin{assumption}\label{a3}
$E(A_t|S_t=s)$, $E(A_t^2|S_t=s)$, $E(A_t^4|S_t=s)$, $E(Y_{t+k}|S_t=s)$, $E(Y_{t+k}^2|S_t=s)$, $p_{S_t}(s)$ as functions of $s$ are uniformly bounded for $s\in \mathcal S$. The first derivatives of these functions are also uniformly bounded.
\end{assumption}
Assumption \ref{a1} is to ensure that the kernel estimators $\hat B_t(s)$, $\hat C_t(s)$, $\hat D_t(s)$ do not diverge to infinity because of $\hat p_{S_t}(s)=\frac{1}{n}\sum_{i=1}^n K_\Lambda(s-S_t^i)$, which converges in probability to $p_{S_t}(s)$, on the denominator. The first equation in Assumption \ref{a2} ensures the unbiasedness of the kernel estimator. When $\mathcal S=\mathbb R^d$, this assumption is satisfied by most commonly used kernel functions. However, when $\mathcal S$ is bounded, a kernel function defined on $\mathbb R^d$ might fail to satisfy this assumption. The rest three equations ensure that the limit distributions of the kernel estimators exist. Assumption \ref{a3} ensures that the higher order terms of the Taylor expansion of the kernel estimators converge to zero. Then we have the following theorem.
\begin{Theorem}\label{normality}
If assumptions \ref{a1}--\ref{a3} are satisfied, and $\Lambda$ satisfies $n|\Lambda|\to \infty$ and $\Lambda\to 0$ as $n\to\infty$, then $\sqrt{n}(\hat \phi_k-\phi_k^*)$ converges to a normal distribution with mean $0$ and variance:
\begin{align*}
    E^{-1}\Big\{H;L_1 (B,C)\Big\} \Sigma(H;\phi_k^*,B,C,D)E^{-1}\Big\{L_1 (H;B,C)\Big\},
\end{align*}
where,
\begin{align*}
\Sigma(H; \phi_k^*, B, C, D)=
\text{Var}\Big\{\mathbb P_n  L_1(H; B, C)\phi_k^*-\mathbb P_n L_2( H; B, C, D)\Big\}.
\end{align*}
 The variance covariance function above can be estimated consistently with:\\
 \begin{align*}
     \mathbb P_n^{-1} \Big\{L_1 (H;\hat B,\hat C)\Big\} \Sigma(H;\phi_k^*,\hat B,\hat C,\hat D)\mathbb P_n^{-1}\Big\{L_1 (H;\hat B,\hat C)]\Big\}.
 \end{align*}
\end{Theorem}
The proof of the theorem is in the appendix. 

\section{Simulation Studies}
We evaluate the proposed method using a simulation study. The following generative model simulates an observational study where the treatment at each time point is correlated with past treatments and covariates. For each individual, data $(X_1,A_1,...,X_{T+1},Y_{T+1})$ are generated as follows: $X_1\sim$ $\text{Normal}(0,\sigma^2)$, $A_1\sim$ $\text{Uniform}(0,1)$; For $t\ge 1$,  $X_{t+1}\sim$ $\text{Normal}(\eta_1 X_t +\eta_2 A_t,$ $\sigma^2)$, $A_{t+1}\sim$ $\text{Normal}(\tau _1 X_{t+1}+\tau_2 A_{t},$ $\sigma^2)$; $Y_{t+1}=$ $\theta_1 X_{t}$ $+\theta_2 A_{t-1}-A_{t}(A_{t}$ $-\beta_{0}-$ $\beta_{1}X_{t})+\epsilon_{t+1}$, where
$\epsilon_t\sim$ $\text{Normal}(0,\sigma^2)$ and the correlation between $\epsilon_{t_1}$ and $\epsilon_{t_2}$ for any $t_1,t_2\in\{2,\dots,T+1\}$ is $\sigma^{|t_1-t_2|/2}$. Here we assume that the data is observed starting from $t=1$ and  the dosages have been transformed so that $A_t\in \mathcal A=\mathbb R$. 

Notice that when $S_t=X_t$ and $\theta_2=0$, assumption (\ref{assumption:strong}) is satisfied (Proof is provided in the appendix). Under the simulation setting above, the true value for the lag $1$ treatment effect is: $\tau_{t,1}(a,S_t)= -a^2 +(\beta_{0}+\beta_{1}S_{t})a$.
We can also prove that for $k\ge 2$, the lag $k$ effect under this generative model also follows a quadratic form 
(See appendix for details). We take $\sigma=0.5$, $\theta_1=0.8$, $\theta_2=0$, $\eta_1=-0.2$, $\eta_2=0.2$, $\tau_1=1$, $\tau_2=-0.5$, $\beta_0=0$, $\beta_1=2$ and $S_t=X_t$.  The true parameters for the lag $1$, lag $2$, lag $3$ treatment effects can thus be calculated: $(\alpha_1,\beta_{1,0},\beta_{1,1})=(-1,0,2)$; $(\alpha_2,\beta_{2,0},\beta_{2,1})=(-0.21,0.16,-0.08)$; $(\alpha_3,\beta_{3,0},\beta_{3,1})= (-0.0125,-0.08,-0.03)$. The true parameters for the lag $3$ weighted advantage with $w_1=w_2=w_3=1/3$ are $(\tilde\alpha_3,\tilde\beta_{3,0},\tilde\beta_{3,1})=(-0.4075,0.0267,0.63)$.  

We generate the dataset with $T=50$ and sample size $n=100,200,400$. We take $S_t=X_t$ and use the proposed method to estimate the treatment effects for lag $1,2$ and $3$. We use the Gaussian kernel $K_{\Lambda}(s)=(2\pi)^{-q/2}|\Lambda|^{-1/2}\exp(-s^T \Lambda s/2)$, where $q=1$ is the dimension of $S_t$, and $f(S_t)=S_t$. In practice, different kernels can be used, which usually will lead to similar results. Here we chose the Gaussian kernel over the others mainly for its computational simplicity. $\Lambda$ is a $q\times q$ diagonal matrix with $\Lambda_{j,j}=\lambda_j^2$. We take $\lambda_j=0.305\times n^{-1/3} \text{sd}(S_{t,j})$, $j=1,\dots,q$. The simulation is replicated for $200$ times with each sample size \footnote{The R code for the simulation can be found in \url{https://github.com/lz2379/Mhealth}.}. The results are presented in Table \ref{table1-sim}.

\begin{table}
\small \renewcommand{\arraystretch}{1}
\begin{center}
 \caption{Simulation results from $200$ replicates for observational studies.\label{table1-sim}}
\begin{threeparttable}
 \begin{tabular}{c| c| c c c c| c c c c| c c c c}
 \toprule
    \multicolumn{2}{c|}{}& \multicolumn{4}{c|}{$\alpha_k$}& \multicolumn{4}{c|}{$\beta_{k,0}$}& \multicolumn{4}{c}{$\beta_{k,1}$}\\
\midrule 
$k$    & n   & Bias$^1$ & SD$^1$ & SE$^1$ & CP & Bias$^1$  & SD$^1$    & SE$^1$ & CP & Bias$^1$  & SD$^1$    & SE$^1$ & CP\\
\midrule
 1       & 100 & 0.9 & 16.5 & 14.8& 93.0     & 0.9 & 9.3 & 9.3  & 95.0     & 1.1  & 39.6 & 35.0 & 95.0   \\
         & 200 & -0.4& 11.1 & 10.4& 93.5     & 0.3 & 6.8 & 6.3 & 91.0      & -0.3 & 25.7 & 24.0 & 92.0 \\
         & 400 & 0.4 & 7.9  & 7.4 & 91.5     & -0.1 & 4.5 & 4.3 & 92.0     & -0.1 & 18.3 & 16.7 & 93.5 \\
\midrule
 2       & 100 & 1.7 &31.6 & 29.0 & 92.5     & -1.0 & 23.0 & 22.3 & 93.0   & -3.1  & 79.7 & 67.9  & 92.0 \\
         & 200 & 0.2 &23.5 & 20.8 & 91.5     & -0.3 & 16.5 & 15.7 & 93.5   & -0.6  & 54.7 & 47.6  & 92.0 \\
         & 400 & -1.8&14.5 & 14.7 & 95.5     & 0.7 & 11.8  & 11.1 & 92.5   & -1.2  & 33.6 & 33.0  & 94.5 \\
 \midrule
 3       & 100 & 2.0  &32.2  & 26.9 & 88.5     &  4.1  & 22.1 &  21.0 & 93.5&  3.1 & 74.8 & 67.3  & 91.0    \\
         & 200 & -3.3 &19.6  & 18.9 & 94.5     &  0.6  & 15.3 &  14.6 & 92.0&  5.8 & 50.5 & 45.6  & 91.5   \\
         & 400 & 1.0  &15.9  & 13.4 & 89.5     &  0.7  & 10.6 &  10.2 & 93.0&  -2.2 & 36.8 & 31.6  & 91.0   \\
\bottomrule
\end{tabular}
\begin{tablenotes}
\footnotesize{
\item[1] Note: These columns are in $10^{-3}$ scale
\item[2] Note: SD refers to the standard deviation of the estimated parameters from $200$ replicates, SE refers to the mean of the estimated standard errors calculated by our covariance function, CP refers to the coverage probability of the $95\%$ confidence intervals calculated using the estimated standard errors.}
\item[3] Note: The worst case Monte Carlo standard error for proportions is $2.3\%$.
\end{tablenotes}
\end{threeparttable}
\end{center}
\end{table}

\begin{table}
    \centering\small
    \caption{Estimated Parameters for Lag 3 Weighted Advantage from 200 Replicates}
    \label{sim-weighted}
    \vskip 0.15in
    \begin{threeparttable}
    \begin{tabular}{c c  c c c}
    \toprule
         n  & $\tilde \alpha_3$    & $\tilde \beta_{3,0}$   & $\tilde \beta_{3,1}$  & $\bar{\tilde \tau}_{K}(\hat\pi_t^{opt},S_t)$ \\
           & ($\times 10^{-2}$)  & ($\times 10^{-2}$)  & ($\times 10^{-2}$) &  ($\times 10^{-3}$) \\
         
         \midrule
         100& -40.6 (2.1)& 2.8 (1.1)& 63.0 (4.7) & 64.7 (0.27)\\
         200& -40.9 (1.4)& 2.7 (0.8) & 63.3 (3.2) & 64.8 (0.13)\\
         400& -40.7 (1.1)& 2.7 (0.5) & 62.9 (2.3) & 64.9 (0.06)\\
        \bottomrule
    \end{tabular}
    \begin{tablenotes}
    \footnotesize{
    \item[1] Note: The numbers in the parenthesis are the standard deviations.
    }
    \end{tablenotes}
    \end{threeparttable}
    \vskip -0.1in
\end{table}


As presented in Table \ref{table1-sim}, the proposed method was able to estimate the parameters with small bias. The standard deviation of the estimated parameters decreased with the sample size increasing. The standard errors estimated with our covariance function provided a close estimate of the standard deviation. The $95\%$ confidence intervals provided a coverage of the true parameters close to $95\%$ in most scenarios. However, the estimated standard errors slightly underestimated the standard deviation, leading to an under-coverage for the confidence intervals. From the proof of Theorem \ref{normality} in the appendix, we see that the variance of the estimated parameters consist of two parts, the variance from the estimation equation and the variance from the kernel estimation. The latter part of the variance converges to $0$ as $n$ goes to infinity and is thus excluded from the asymptotic variance formula. However, when the sample size is not large enough, excluding this part of the variance might lead to underestimation of the variance, as supported by the simulation result. 

Table \ref{sim-weighted} presents the estimated parameters for the lag $3$ weighted advantage with $w_1=w_2=w_3=1/3$ from 200 replicates for each sample size. For each replicate, we obtain $\hat \pi_t^{opt}=-\hat{\tilde\beta}_K^TS_t/(2\hat{\tilde\alpha}_K)$ and calculate the lag $3$ weighted advantage of this suggested treatment regime. The lag $3$ weighted advantage is calculated on a test dataset with $5000$ subjects each with observations from time $t=1,\dots,T+3$. 
Table \ref{sim-weighted} presents the average lag $3$ weighted advantage across time $\bar{\hat\tau}_K=\sum_{t=1}^T\tilde \tau_{t,K}(a=\hat\pi_t^{opt},S_t)/T$. The average lag $3$ weighted advantage of the true optimal treatment regime 
is $65.0\times 10^{-3}$. As the result shows, the treatment regimes estimated by the proposed method was close to optimal.

In order to see how the model performs when assumption (\ref{assumption:strong}) is not satisfied, we generate the datasets with the same parameters except that $\theta_2=-0.1$. Under this setting, assumption (\ref{assumption:strong}) is not satisfied for $k=1$ when $S_t=X_t$ (see appendix for details). 
The result of the simulation is presented in Table \ref{table-sim-biased}.  The estimated parameters for $\beta_{1,0}$ were biased, thus leading to wrong statistical inference of the parameters. Since for $k=2,3$, assumption (\ref{assumption:strong}) is still satisfied, the result remained unbiased (see appendix for the complete results). 
We also calculate the average lag 3 weighted advantage of the treatment regime suggested by the biased estimation equation with $w=(1/3,1/3,1/3)$. The average lag 3 weighted advantage of the true optimal treatment regime is $64.5\times 10^{-3}$, while the average lag 3 weighted advantages of the estimated treatment regime are $63.9\times 10^{-3}$, $64.0\times 10^{-3}$ and $64.1\times 10^{-3}$ for sample size $100$, $200$ and $400$. 
In this particular setting, the recommended treatment regime was still close to optimal. However, it cannot be guaranteed that the suggested treatment regime would be close to optimal in a different setting. One solution to the bias is to include more information in $S_t$. Under this specific setting, it is trivial to prove that $Y_{t+k}, A_t|X_t, A_{t-1}$. Therefore, by taking $S_t=(X_t,A_{t-1})$, we could obtain unbiased estimates of the parameters using the same estimation equation. The estimated results with $S_t=(X_t,A_{t-1})$ are given in the appendix. 

\begin{table}[t]
    \centering
    \caption{Estimated Parameters for Lag 3 Weighted Advantage from 200 Replicates When $\theta_2=-0.1$}\label{table-sim-biased}
    \begin{threeparttable}
    \begin{tabular}{c c  c c c}
    \toprule
         n  & $\tilde \alpha_3$   & $\tilde \beta_{3,0}$  & $\tilde \beta_{3,1}$ & $\tilde \tau_{K}(\hat\pi_t^{opt},S_t)$ \\
         \midrule
         100& -40.6 (2.1)& 2.9 (1.1)& 62.9(4.7) & 63.9(0.46)\\
         200& -40.9 (1.4)& 2.7 (0.8) & 63.3(3.2) & 64.0(0.25)\\
         400& -40.7 (1.0)& 2.8 (0.5) & 62.9(2.3) & 64.1(0.17)\\
    \bottomrule
    \end{tabular}
    \begin{tablenotes}
    \footnotesize{
    \item[1] Note: Columns 2-4 are in $10^{-2}$ scale; column 5 is in $10^{-3}$ scale.
    \item[2] Note: The numbers in the parenthesis are the standard deviations.
    \item[3] Note: The last column $\tilde \tau_{K}(\hat\pi_t^{opt},S_t)= \sum_{t=1}^T\tilde \tau_{t,K}(a=\hat\pi_t^{opt},S_t)/T$.}
    \end{tablenotes}
    \end{threeparttable}
\end{table}

\section{Type 1 Diabetes Data Analysis }
Rapid-reacting insulin therapies are frequently used for diabetes patients before meals to prevent hyperglycemia events. However, the patients under the insulin therapies may be constantly under the risk of hypoglycemia or hyperglycemia due to overdosing or underdosing. Mobile technologies can provide real-time tracking on blood glucose, physical activity and insulin injections of the patients and thus facilitate the dose adjustments to prevent adverse events \citep{maahs2012outpatient}. 
We apply our method to the Ohio type 1 diabetes dataset collected by \citet{marling2018ohiot1dm} to estimate the lagged treatment effects of the doses and then provide dose suggestions which maximize the weighted advantage\footnote{The R code for real data application can be found in \url{https://github.com/lz2379/Mhealth}.}. 

This dataset contains six patients, each with eight weeks of data, including: blood glucose; insulin dosages, including rapid reacting insulin taken before meals (bolus insulin doses), and long-term insulin infused continuously through out the day (basal insulin doses); sensor-collected physiological measurements including heart rate, body temperature and steps; and self-reported life-events including carbonhydrates intake and exercises. Through exploratory analysis, we found that each patient has distinct patterns in insulin usage and blood glucose levels. Therefore, we regard them as 6 separate datasets. We illustrate with the data of one patient and assume that the data from each day of this patient are independent from each other. Results of the other patients are presented in the appendix. There are $54$ days of data available. We further take the first $44$ days as the training data and the last $10$ days as the testing data. 
We summarize the measurements every $30$ minutes, resulting in $T=48$ time intervals each day.  For each $30$-minute time interval, the covariates we consider include total carbonhydrates intake, planned total carbonhydrates intake in the next time interval, average glucose level, average heart rate and basal insulin level. We denote these covariates as: $X_t=(\text{Carb}_t$, $\text{Carb-Planned}_t$, $\text{Glucose}_t$,$\text{Heartrate}_t$,$\text{Basal}_t)^T$.  Since education of meal planning is typically incorporated as a part of the insulin therapy for diabetes patients \citep{bantle2008nutrition}, we assume that all the carbonhydrates intake within $30$ minutes are planned ahead of time and 
$\text{Carb-Planned}_{t}=\text{Carb}_{t+1}$. $A_t$ is the total bolus injection from $t-1$ to $t$. Let $A_{\text{max}}$ be the maximum observed dose across all days and time. We estimate $\mathcal A$ with the interval $[0, A_{\text{max}}]$. $Y_t$ is taken to be the average of the index of glycemic control (IGC) between time $t-1$ and $t$ calculated by:
\begin{align*}
    \text{IGC}=
    -\frac{I(\text{G}<80) |80-\text{G}|^{2}}{30}-
    \frac{I(\text{G}>140)|\text{G}-140|^{1.35}}{30}
\end{align*}
where $G$ is the measured blood glucose level (See \citet{rodbard2009interpretation} for various criterias for glycemic control evaluation). Higher $Y_t$ indicates a better glycemic control within the time interval. We take $S_t=(X_t^T$, $\text{Basal-4-8-hour}_t$, $A_{t-1})^T$, where $\text{Basal-4-8-hour}_t$ $=\sum_{l=8}^{15}\text{Basal}_{t-l}/8$. These covariates are chosen because they are significantly correlated with $A_t$ from exploratory analysis. To satisfy assumption (\ref{assumption:strong}), all covariates correlated with $A_t$ need to be included in $S_t$. We take $f(S_t)=$ $(\text{Carb}_t,$ $\text{Carb-Planned}_t,$ $\sum_{k=8}^{15}\text{Basal}_{t-k}/8,$ $A_{t-1})$ and predict the treatment effect of the dosage within two hours, $k=1, \dots, 4$. Thus the model for the lag $k$ causal effect can be written as:
\begin{align*}
&\tau_k(a,S_t)=\alpha_k a^2+(\beta_{k,0}+\beta_{k,1}\text{Carb}_t+\beta_{k,2}\text{Carb-Planned}_t+\beta_{k,3}\text{Basal-4-8-hour}_t+\beta_{k,4}A_{t-1})a    
\end{align*}
We still use Gaussian kernel and the bandwidth $\Lambda$ is chosen to be a $q\times q$ diagonal matrix with $\Lambda_{j,j}=\lambda_j^2$ and $\lambda_j=0.305\times n^{-1/8}\text{sd}(S_{t,j})$, where $ j=1,...,q$ and $q=7.$


\begin{table}[t]
\centering
 \caption{Estimated variables with the Ohio type 1 diabetes dataset\label{table2-realdata}}
\begin{threeparttable}
 \begin{tabular}{c| c c c c c}
    \toprule
    $k$ &1&2&3&4&Weighted\\
    \midrule
    $\alpha_k$ ($\times 10^{-2}$)    &-12.7(9.0) &-20.6(14.8) &-13.7(12.8) &-5.2(12.1) & -13.0(11.0)  \\
    $\beta_{k,0}$ ($\times 10^{-1}$) &15.8(8.2)  &45.6(14.7)  &50.0(11.1)  &33.8(17.8) & 36.9 (10.9)  \\
     $\beta_{k,1}$ ($\times 10^{-3}$)&21.8(10.8) &17.4(15.5)  &15.3(18.7)  &25.3(20.3) & 15.2 (13.7)\\
     $\beta_{k,2}$ ($\times 10^{-3}$)&25.1(10.1) &18.8(13.8)  &8.8(14.5)   &6.5(16.5) & 13.8(11.9)\\
     $\beta_{k,3}$ ($\times 10^{-1}$)&-15.6(9.7) &-40.5(14.8) &-47.4(12.5) &-35.2(18.7)&34.9(11.8)\\
     $\beta_{k,4}$ ($\times 10^{-2}$)&-6.1(11.6) &-15.4(19.6) &-8.2(24.9)  &-14.3(25.8)&-9.1(18.3)\\
 \bottomrule
\end{tabular}
\begin{tablenotes}
\footnotesize{
\item[1] Note: The numbers in the parenthesis are the estimated standard errors calculated by the covariance formula.
\item[2] Note: The last column presents the estimated parameters for the lag $4$ weighted advantage with $w_1=w_2=w_3=w_4=1/4$.
}
\end{tablenotes}
\end{threeparttable}
\end{table}

The estimated parameters are presented in Table \ref{table2-realdata}. 
The optimal treatment would be the one which maximizes the weighted advantage for two hours. Since the estimated $\tilde \alpha_K$ was negative, the optimal treatment regime at time $t$ can be estimated by $\hat \pi_t^{opt}=$ $-\{\hat {\tilde\beta}_K^T S_t\}/(2\hat{\tilde \alpha}_K)$ when $-\{\hat {\tilde\beta}_K^T S_t\}/(2\hat{\tilde \alpha}_K)$ $\in [0,A_{\text{max}}]$; $0$ when $-\{\hat {\tilde\beta}_K^T S_t\}/(2\hat{\tilde \alpha}_K)$ $<0$; $A_{\text{max}}$ when $-\{\hat {\tilde\beta}_K^T S_t\}/(2\hat{\tilde \alpha}_K)>A_{\text{max}}$. Since $\hat{\tilde\alpha}_K<0$, the parameters $\hat{\tilde\beta}_{K,j}$ can be interpreted as the units of increase in optimal insulin with $-2\tilde\alpha_K$ extra units in the $S_{t,j}$ had the other covariates held constant. 
The results implied that the optimal dose should be higher when the carbonhydrates intake was higher over the past half an hour or the planned carbonhydrates intake is higher for the next half an hour ($\hat{\tilde \beta}_{K,1},\hat{\tilde\beta}_{K,2}>0$); the optimal dose should be lower when the average basal insulin rate $4$ to $8$ hours ago was higher or the dose in the last half an hour was higher ($\hat{\tilde \beta}_{K,3},\hat{\tilde\beta}_{K,4}<0$). These results are consistent with the fact that carbonhydrates intake increases the blood glucose and past insulin injections lower the blood glucose. The result also implies that the past basal insulin infusion rate is an important factor in deciding the optimal insulin dosage for the current moment. 

We then estimate the lag $K$ weighted advantage on the test dataset using the estimated parameters $\hat {\tilde \tau}_{t,K}(a,S_t)=\hat{\tilde\alpha}_K a^2+\hat{\tilde\beta}_K f(S_t)$. The mean of the estimated average lag $4$ weighted advantage $\sum_{t=1}^T{\tilde \tau}_{t,K}(a,S_t)/T$ is $0.63$ for the suggested treatment regime and $0.13$ for the original doses. If the model was correct, this method could be used to provide dose suggestions which enhance the stability of the blood glucose for diabetes patients within two hours. 

\section{Discussion and Conclusion}

In this article, we defined the lag $k$ treatment effects for continuous treatments following the framework by \citet{boruvka2018assessing}. Nonparametric structural nested models with a quadratic form were used for estimating the causal effects of continuous treatments based on mobile health data. We also defined the weighted lag $K$ advantage to measure the effect of the treatments within a short time period in the future. The optimal treatment regime was defined to be the one which maximizes this advantage. The R code for the simulations and the real data application is provided in the supplementary material.

The proposed method fills the gap in the literature of sequential decision making where the goal is to provide dose suggestions which maximize short-term outcomes. This semiparametric model provides more robustness against model misspecification. By conditioning on partial information of the past history, the proposed method excludes irrelevant information for the estimation of the optimal treatment regime. Thus, the complexity of the suggested optimal dosage would not increase as $T$ increases and is more practical when applied to infinite horizon data. Compared to other infinite horizon methods where stationarity or Markovian property is required, this method imposes minimal assumptions on the data generating process. Statistical inference can also be provided for the estimated parameters. The simulation studies showed that the method was capable of estimating the parameters accurately and the variance could be approximated with our covariance function. The estimated treatment regime was close to maximizing the weighted lag $K$ advantage. Application to the Ohio type 1 diabetes dataset showed that this method could provide meaningful insights for dose suggestions based on observed history of the patients. In practice, to ensure the unbiasedness of proposed estimation equation, it is essential to include all confounders which influence both $A_t$ and $Y_{t+k}$ into $S_t$.

The proposed method is also subject to a few limitations. 
 First,  the proposed method is limited to estimating the causal effect of a one-time change in the treatment history. Estimating the cumulative treatment effects if all future treatments follow the suggested treatment regime would be of more interest in certain scenarios. However, the single-time change measurement can still be of great use in practice. In real life, when patients are taking medications, they are likely to take medications only a few times per day. It would be useful to measure what would be the best dosage if the patient takes the medication at the moment and conduct no further medical actions for the next few hours.
 Second, when the key assumption (\ref{assumption:strong}) is not satisfied, the proposed method might lead to biased results. In Appendix C.2, we showed that this bias can be avoided by including all variables that are influential to the decision making process. In practice, for diseases like diabetes and hypertension (which are the main applications we are interested in), patients typically receive education from clinicians on dosage calculation before starting the treatment. Take diabetes as an example. Key decision-making factors, including meals, exercise levels, physical indicators, are collected by most blood glucose monitoring applications. The proposed method can be applied to enhance the performance of the dosages when the key factors for dosing are well established for the patients and can be easily collected. However, the assumption (\ref{assumption:strong}) might be hard to examine when the decision making process of the patient is unknown. Therefore, it is definitely essential to let patients be aware of the possible fallacy of the method when an outside factor is guiding his/her decision making process.
 Third, due to the quadratic form of the model for the lagged treatment effects, a slight underestimate of the quadratic effects of the doses may lead to a large overestimate of the optimal dose. Possible future work would include improving the method to avoid overestimation in doses. 
 
 In the future, we are also interested in extending the method to incorporate both long-term optimization and short-term monitoring. Applying this method to online streaming data is also of great interest. For a fixed group of users, the proposed method can be extended to allow streaming data without much additional computation. If the number of users is large, kernel estimation would be computationally expensive. One potential solution is to divide users into subgroups according to certain demographic or medical similarities and then conduct kernel estimation within each subgroup. 
 
 
\nocite{*}
\bibliography{paperref}
\bibliographystyle{icml2020}

\appendix

\subsection{Proof of Equation (\ref{potential-observed})}
\label{section:mhealth-proof-potential}
First, for the first term in equation (\ref{potential-observed}), we can derive:
\begin{align*}
    &E\Big[Y_{t+k}(\bar A_{t-1} , a , A_{t+1}^{a_t=a} , \dots, A_{t+k-1}^{a_t=a} )|H_t(\bar A_{t-1})\Big]\\
    &=E\Big[Y_{t+k}(\bar A_{t-1} , a , A_{t+1}^{a_t=a} , \dots,  A_{t+k-1}^{a_t=a} )|H_t(\bar A_{t-1}), A_t=a\Big]\\
    &=E\Big[Y_{t+k}(\bar A_{t-1} , A_t , A_{t+1}^{a_t=a} , \dots, A_{t+k-1}^{a_t=a} )|H_t(\bar A_{t-1}), A_t=a\Big]\\
    &=E\Big[E\Big\{Y_{t+k}(\bar A_{t-1} , A_t , A_{t+1}^{a_t=a} , \dots, A_{t+k-1}^{a_t=a} )|H_t(\bar A_{t-1}),A_t=a,A_{t+1}=A_{t+1}^{a_t=a},\dots,\\& A_{t+k-1}=A_{t+k-1}^{a_t=a}\Big\}|H_t(\bar A_{t-1}), A_t=a \Big]\\
    &=E\Big[E\Big\{Y_{t+k}|H_t(\bar A_{t-1}),A_t=a,A_{t+1}=A_{t+1}^{a_t=a},\dots, A_{t+k-1}=A_{t+k-1}^{a_t=a}\Big\}|H_t(\bar A_{t-1}), A_t=a \Big]\\
    &=E\Big[Y_{t+k}|H_t(\bar A_{t-1}), A_t=a\Big]\\
    &=E\Big[Y_{t+k}|H_t, A_t=a\Big],
\end{align*}

where the first equation is based on the sequential ignorability assumption; The second, the third and the fifth equations are based on the property of the conditional expectation; The fourth and the last equations are based on the consistency assumption. 
Equation (\ref{potential-observed}) can thus be proved.

\subsection{Proof of Equation (\ref{unbiasedness})}
We need to show that under assumptions (\ref{model}) and (\ref{assumption:strong}) :
\begin{equation*}
    E\Bigg[\Big\{d(A_t,S_t)-E(d(A_t,S_t)|S_t)\Big\}\times \Big\{U_{t+k}-E(U_{t+k}|S_t)\Big\}\Bigg]=0.
\end{equation*}
By the property of conditional expectation, it is trivial to obtain that: 
\begin{align*}
    &E\Big[\Big\{d(A_t,S_t)-E(d(A_t,S_t)|S_t)\Big\}\times E(U_{t+k}|S_t)\Big]\\
    &=E\Bigg(E\Bigg[\Big\{d(A_t,S_t)-E(d(A_t,S_t)|S_t)\Big\}\times E(U_{t+k}|S_t)\Bigg|S_t\Bigg]\Bigg)\\
    &=E\Big[E\Big\{d(A_t,S_t)|S_t\Big\}-E\Big\{d(A_t,S_t)|S_t)\Big\}\times E(U_{t+k}|S_t)\Big]\\
    &=0.
\end{align*}
Thus Equation (\ref{unbiasedness}) is equivalent to:
$
    E\Big[\Big\{d(A_t,S_t)-E(d(A_t,S_t)|S_t)\Big\}\times U_{t+k}\Big]=0.
$
By the property of conditional expectation, it is sufficient to show that:
\begin{align*}
    E\Big[\Big\{d(A_t,S_t)-E(d(A_t,S_t)|S_t)\Big\}\times U_{t+k}\Big |S_t\Big]=0,
\end{align*}
which is equivalent to:
\begin{align}\label{equation:snmm2}
    E\Big[d(A_t,S_t)U_{t+k}|S_t\Big]=E\Big[d(A_t,S_t)|S_t\Big]E\Big[U_{t+k}|S_t\Big].
\end{align}
From the definition of $U_{t+k}$, we can obtain that:
\begin{align*}
    &U_{t+k}=Y_{t+k}-\tau_{k}(A_t,a_0,S_t).
\end{align*}
With consistency assumption, $Y_{t+k}=Y_{t+k}(\bar A_{t-1},a=A_t, A_{t+1}^{a=A_t},\dots,A_{t+k-1}^{a=A_t})$. Thus,
\small
\begin{align*}
    &U_{t+k}=Y_{t+k}(\bar A_{t-1},a_t=A_t, A_{t+1}^{a_t=A_t},\dots,A_{t+k-1}^{a_t=A_t}) -\tau_{k}(A_t,a_0,S_t).
\end{align*}
\normalsize
By the consistency assumption , it is trivial to prove that $S_t(\bar A_{t-1})=S_t$. Then,
\small
\begin{align*}
    &E(U_{t+k}|S_t,A_t)=E\Bigg[Y_{t+k}(\bar A_{t-1},a_t=A_t, A_{t+1}^{a_t=A_t},\dots,A_{t+k-1}^{a_t=A_t}) - \tau_{k}(A_t,a_0,S_t)|S_t(\bar A_{t-1}),A_t\Bigg]\\
    &=E\Bigg[Y_{t+k}(\bar A_{t-1},a_t=A_t, A_{t+1}^{a_t=A_t},\dots,A_{t+k-1}^{a_t=A_t}) -
    E\Big\{Y_{t+k}(\bar A_{t-1},a_t=A_t, A_{t+1}^{a_t=A_t},\dots,A_{t+k-1}^{a_t=A_t})- \\&Y_{t+k}(\bar A_{t-1},a_t=a_0, A_{t+1}^{a_t=a_0},\dots,A_{t+k-1}^{a_t=a_0})\Big|H_t(\bar A_{t-1}),A_t\Big\}\Big|S_t(\bar A_{t-1}),A_t\Bigg].
\end{align*}
\normalsize

We first take the conditional expectation with respect to $H_t(\bar A_{t-1}), A_t$. Then the first term and the second term are both $$E\{Y_{t+k}(\bar A_{t-1},a_t=A_t, A_{t+1}^{a_t=A_t},\dots,A_{t+k-1}^{a_t=A_t})|H_t(\bar A_{t-1}),A_t\}$$ and can be canceled. Thus the right side of the above equation is equal to:
\begin{align*}
    &E\Big[E\{Y_{t+k}(\bar A_{t-1},a_t=a_0, A_{t+1}^{a_t=a_0},\dots,A_{t+k-1}^{a_t=a_0})|H_t(\bar A_{t-1}),A_t\}|S_t(\bar A_{t-1}),A_t\Big]\\
    &=E\Big[Y_{t+k}(\bar A_{t-1},a_t=a_0, A_{t+1}^{a_t=a_0},\dots,A_{t+k-1}^{a_t=a_0})|S_t(\bar A_{t-1}),A_t\Big]\\
    &=E\Big[Y_{t+k}(\bar A_{t-1},a_t=a_0, A_{t+1}^{a_t=a_0},\dots,A_{t+k-1}^{a_t=a_0})|S_t(\bar A_{t-1})\Big],
\end{align*}
where the last equation is based on assumption (\ref{assumption:strong}). Therefore, \begin{align*}
    E(U_{t+k}|S_t,A_t)=E\Big[Y_{t+k}(\bar A_{t-1},a_t=a_0, A_{t+1}^{a_t=a_0},\dots,A_{t+k-1}^{a_t=a_0})|S_t(\bar A_{t-1})\Big].
\end{align*}
Take expectation with respect to $S_t$ for both sides, we obtain that:
\begin{align*}
    E(U_{t+k}|S_t)=E\Big[Y_{t+k}(\bar A_{t-1},a_t=a_0, A_{t+1}^{a_t=a_0},\dots,A_{t+k-1}^{a_t=a_0})|S_t(\bar A_{t-1})\Big]=E(U_{t+k}|S_t,A_t).
\end{align*}
Therefore, 
\begin{align*}
    &E\Big[d(A_t,S_t)U_{t+k}|S_t\Big]\\
    &=E\Big[E\{d(A_t,S_t)U_{t+k}|S_t,A_t\}\Big |S_t\Big]\\
    &=E\Big[  d(A_t,S_t) E\{U_{t+k}|S_t,A_t\}\Big|S_t\Big]\\
    &=E\Big[  d(A_t,S_t) E\{U_{t+k}|S_t\}\Big|S_t\Big]\\
    &=E\Big\{  d(A_t,S_t)|S_t\Big\} E\Big\{U_{t+k}|S_t\Big\}.
\end{align*}
Thus Equation (\ref{equation:snmm2}) can be proved. Therefore, Equation (\ref{unbiasedness}) is proved.  

\subsection{Proof of Theorem \ref{normality}}
First of all,
\begin{align*}
    &\sqrt{n} (\hat \phi_k-\phi_k^*)\\
    =&\Bigg[\mathbb P_n L_1(H;\hat B,\hat C)\Bigg]^{-1}\Bigg[ \sqrt{n} \mathbb P_n\Big\{ L_2(H;\hat B,\hat C,\hat D)- L_1(H;\hat B,\hat C)\phi_k^* \Big\}\Bigg].
\end{align*}

The second part on the right side can be written as:
\begin{align*}
    & \sqrt{n}\mathbb P_n  \Big\{ L_2(H;\hat B,\hat C,\hat D)-L_1(H;\hat B,\hat C)\phi_k^* \Big\}\\
    =&\sqrt{n} \Big[\mathbb P_n\{ L_2(H;\hat B,\hat C,\hat D)- L_1(H;\hat B,\hat C)\phi_k^*\} -
    \mathbb P_n \{L_2(H; B, C, D)- L_1(H; B, C)\phi^*_k\}\Big]
    \\&+\sqrt{n} \Big[ \mathbb P_n \{L_2(H;B, C, D)-L_1( H;B, C)\phi^*_k\} -
    E\{L_2(H;B,C,D)-L_1(H;B,C)\phi^*_k\}\Big].
\end{align*}

Therefore, to prove Theorem \ref{normality}, it is enough to show the following three equations:
\begin{align}
    &\mathbb P_n L_1(H;\hat B,\hat C) \xrightarrow{p} E[L_1(H;B,C)],\label{eq1}\\
    \sqrt{n} \Big[&\mathbb P_n\{ L_2(H;\hat B,\hat C,\hat D)- L_1(H;\hat B,\hat C)\phi_k^*\} -\mathbb P_n \{L_2( H;B, C, D)-L_1( H;B, C)\phi^*_k\}\Big]\nonumber\\&=o_p(1),\label{eq2}\\
    \sqrt{n} \Big[& \mathbb P_n\{ L_1( H;B, C)\phi_k^* -L_2( H;B, C, D)\}-E\{ L_1( H;B, C)\phi_k^*-
   L_2(H; B, C, D)\} \Big]
   \nonumber\\&\xrightarrow{d} N\Big\{0,\Sigma(\phi_k^*;B,C,D)\Big\}.\label{eq3}
\end{align}

Then with Slutsky's theorem, we can obtain that $\sqrt{n}(\hat \phi_k-\phi_k^*)$ converges in distribution to a mean zero normal random vector with variance-covariance matrix given by:
\begin{align*}
    E^{-1}\Big\{L_1(H;B,C)\Big\}\Sigma(H;\phi_k^*, B, C, D)E^{-1}\Big\{L_1(H;B,C)\Big\}.
\end{align*}

\subsubsection{Proof of Equation (\ref{eq1})}\label{section:basic_calc}
First, we can obtain:
\begin{align*}
    &\mathbb P_n L_1(H;\hat B,\hat C) -E [L_1(H;B,C)]\\
    &=\{\mathbb P_n L_1(H;\hat B,\hat C) -\mathbb P_n L_1(H;B,C)\} 
    +\{\mathbb P_n L_1 (H;B,C)-E [L_1(H;B,C)]\}.
\end{align*}

The second part on the right is $o_p(1)$ by the law of large numbers. Therefore, we just need to prove that the first part is $o_p(1)$. With Taylor expansion and the mean value theorem, we can obtain:
\begin{gather*}
    \Big|\mathbb P_n L_1(H;\hat B,\hat C) -\mathbb P_n L_1(H;B,C)\Big|\\
    =\Big|\mathbb P_n \Big\{\frac{\partial L_1(H;B,C)}{\partial B}\Big|_{B'} (\hat B-B)+
    \frac{\partial L_1(H;B,C)}{\partial C}\Big|_{C'} (\hat C-C)\Big\}\Big|\\
    \le \Big|\mathbb P_n\Big\{\frac{\partial L_1(H;B,C)}{\partial B}\Big|_{B'} (\hat B-B)\Big\}\Big|+\Big|\mathbb P_n \Big\{ \frac{\partial L_1(H;B,C)}{\partial C}\Big|_{C'} (\hat C-C)\Big\}\Big|,
\end{gather*}
for some $B'$ between $\hat B$ and $B$, and $C'$ between $\hat C$ and $C$. Since, 
\begin{align}\label{equation:mhealth-L1sup}
    \Big|\mathbb P_n\Big\{\frac{\partial L_1(H;B,C)}{\partial B}\Big|_{B'} (\hat B-B)\Big\}\Big| \le \mathbb P_n \Bigg|\frac{\partial L_1(H;B,C)}{\partial B}\Big|_{B'}\Bigg| \sup_{s\in\mathcal S}\Big|\hat B-B\Big|.
\end{align}

Notice that:
\begin{align*}
    &E\Bigg|\frac{\partial L_1(H;B,C)}{\partial B_t}\Big|_{B'}\Bigg|\\
    &=E\begin{pmatrix}
        2|B_t'-A_t^2| & |C_t-A_t|f_k(S_t)\\
        |C_t-A_t| f_k(S_t)&0
    \end{pmatrix}.
\end{align*}
By assumption \ref{a3}, we can obtain that $E\{|C_t-A_t|f_k(S_t)\}<\infty$. 
Furthermore, $E|B_t'-A_t^2|\le E|B_t'-B_t| +E|B_t-A_t^2| \le E|\hat B_t-B_t| +E|B_t-A_t^2|$. By assumption \ref{a3}, we can obtain that $E\{B_t-A_t^2\}< \infty. $ Thus, 
If we can prove that:
\begin{align}\label{Bkernel}
    &\sup_{s}|\hat B_t(s)-B_t(s)|=o_p(1),
\end{align} 
then $E\Big|\frac{\partial L_1(H;B,C)}{\partial B}|_{B'}\Big|<\infty$. 
Since
$$\mathbb P_n\Bigg|\frac{\partial L_1(H;B,C)}{\partial B}\Big|_{B'}\Bigg| \xrightarrow{p}E\Bigg|\frac{\partial L_1(H;B,C)}{\partial B_t}\Big|_{B'}\Bigg|,$$ 
we obtain that:
$$\mathbb P_n\Bigg|\frac{\partial L_1(H;B,C)}{\partial B}\Big|_{B'}\Bigg|=O_p(1).$$ Together with Equation (\ref{Bkernel}), we obtain that the right side of Equation (\ref{equation:mhealth-L1sup}) is $o_p(1)$.
Similarly, if we can prove that:
\begin{align}\label{Ckernel}
    &\sup_{s}|\hat C_t(s)-C_t(s)|=o_p(1),
\end{align} 
then we can obtain:
$$\mathbb P_n \Bigg|\frac{\partial L_1(H;B,C)}{\partial C}\Big|_{C'} (\hat C-C)\Bigg|=o_p(1).$$
Thus we can obtain that:
\begin{align*}
    \Big|\mathbb P_n L_1(H;\hat B,\hat C) -\mathbb P_n L_1(H;B,C)\Big|=o_p(1).
\end{align*}
Together with $\mathbb P_n L_1(H;B,C)-E[L_1(H;B,C)]\xrightarrow{p}0$ by the law of large numbers, we can finish the proof for equation (\ref{eq1}).

Below, we prove equation (\ref{Ckernel}). Proof of equation (\ref{Bkernel}) can be derived similarly. 
First, let the density of $S_t$ be $f_{S_t}$ and $\hat f_{S_t}(s)=\{\sum_{i=1}^n K_{\lambda}(s-S_t^i)\}/n$. Write $\hat C_t(s)$ as: $\hat C_t(s)=\hat C_{t,1}(s)/\hat f_{S_t}(s)$, where $\hat C_{t,1}(s)=\{\sum_{i=1}^n A_t^i K_{\lambda}(s-S_t^i)\}/n$. Also let $C_{t,1}(s)=C_t(s)f_{S_t}(s)$, then:
\begin{align*}
    &\sup_{s}|\hat C_t(s)-C(s)|=\sup_{s}|\frac{\hat C_{t,1}(s)}{\hat f_{S_t}(s)}-\frac{C_{t,1}(s)}{f_{S_t}(s)}|=\\
    &\sup_{s} \Big|\frac{\big\{\hat C_{t,1}(s)-C_{t,1}(s)\big\}f_{S_t}(s)-C_{t,1}(s)\big\{\hat f_{S_t}(s)- f_{S_t}(s)\big\}}{\hat f_{S_t}(s) f_{S_t}(s)}\Big|\\
    &\leq \sup_{s}\Big|\frac{\hat C_{t,1}(s)-C_{t,1}(s)}{\hat f_{S_t}(s)}\Big|+
    \sup_{s}\Big|\frac{C_{t,1}(s)\big\{\hat f_{S_t}(s)- f_{S_t}(s)\big\}}{\hat f_{S_t}(s) f_{S_t}(s)}\Big|.
\end{align*}

Under the boundedness of $C_{t,1}(s)$ and the assumption that $p_{S_t}(s)$ is uniformly bounded away from $0$, it suffices to show that:
\begin{gather}
    \sup_{s}|\hat C_{t,1}(s)-C_{t,1}(s)|\to 0, \label{C1kernel}\\
    \sup_{s}|\hat f_{S_t}(s)-f_{S_t}(s)|\to 0 \label{pskernel}.
\end{gather}
We demonstrate the  proof for equation (\ref{C1kernel}). Equation (\ref{pskernel}) can be proved similarly. First notice:
\begin{gather}
    \sup_{s}|\hat C_{t,1}(s)-C_{t,1}(s)|\le
    \sup_{s}|\hat C_{t,1}(s)-E\{\hat C_{t,1}(s)\}|+\sup_{s}|E\{C_{t,1}(s)\}-C_{t,1}(s)|. \label{twoparts}
\end{gather}

We prove the uniform convergence of the two parts on the right separately. First, we obtain:
\begin{align*}
    &E\{\hat C_{t,1}(s)\}=E\{A_tK_\Lambda(s-S_t)\}\\
    &=\int_{a_t}\int_{s_t}a_tK_\Lambda(s-s_t)f_{A_t|S_t}(a_t|s_t)f_{S_t} (s_t)ds_t da_t\\
    &=\int_{s_t}C_t(s)|\Lambda^{-1/2}|K\big(\Lambda^{-1/2}(s-s_t)\big)f_{S_t} (s_t)ds_t.
\end{align*}

Let $v=\Lambda^{-1/2}(s-s_t)$, then $s_t=s-\Lambda^{1/2}v$. Let $\mathcal V_s= \{v:s-\Lambda^{-1/2}v\in\mathcal S\}$, then the above equation is equal to:
\begin{align*}
    &\int_{\mathcal V_s}C_t(s-\Lambda^{1/2}v)K(v)f_{S_t}(s-\Lambda^{1/2}v)dv\\
    &=\int_{\mathcal V_s}\Big\{C_t(s)-v^T\Lambda^{1/2}\dot{C}_t(s')\Big\} K(v) \Big\{f_{S_t}(s)-v^T\Lambda^{1/2}\dot{f}_{S_t}(s'')\Big\}dv\\
    &=C_t(s)f_{S_t}(s)\Big\{\int_{\mathcal V_s} K(v)dv\Big\}-\Big\{C_t(s)\dot f_{S_t}(s'')^T+f_{S_t}\dot C_t(s')^T\Big\}\Lambda^{\frac{1}{2}}\Big\{\int_{\mathcal V_s} vK(v)dv\Big\},
\end{align*}

where the first equation above is obtained by Taylor expansion; $s'$ and $s''$ are vectors on the segment connecting $s$ and $S_t$; for any function $g(s)$, $\dot g(s)=\partial g(s)/\partial s$.
From the assumptions, $\Lambda\to 0$ as $n\to\infty$, thus $\inf_s \mathcal V_s\to \mathcal S$. $\inf_s \{\int_{\mathcal V_s} K(v)dv\}=1-O(\Lambda^{1/2})$. $\int_{\mathcal V_s} vK(v) dv\le \int_{\mathcal S} vK(v)=O(1)$. Thus $E\{\hat C_{t,1}(s)\}=C_{t,1}(s) +O(\Lambda^{1/2})$. 
Next, we prove the uniform convergence of the first part of equation (\ref{twoparts}). 
\begin{align*}
    &\sup_s|\hat C_{t,1}(s)-E\{\hat C_{t,1}(s)\}|\\
    =&\sup_s|\frac{1}{n}\{\sum_{i=1}^nA_t^iK_\Lambda(s-S_t^i)\}-E\{ A_t^iK_\Lambda(s-S_t^i)\}|\\
    =&\sup_s\Big|\int_{s_t}C_t(s_t)|\Lambda^{-\frac{1}{2}}|K\Big(\Lambda^{-\frac{1}{2}}(s-s_t)\Big)d\big\{F_n(s_t)-F(s_t)\big\}\Big|,\\
\end{align*}
where $F_n(s_t)$ and $F(s_t)$ denote the empirical cumulative distribution and the cumulative distribution of $S_t$. Then with integration by part, the above equation is less or equal to:
\begin{align*}
    & |\Lambda^{-\frac{1}{2}}| \sup_{s,s_t}\Big|C_t(s_t)K\big(\Lambda^{-\frac{1}{2}}(s-s_t)\big)\Big\{F_n(s_t)-F(s_t)\Big\}\Big|\\
    &+\sup_s\Big|\int_{\mathcal S}\Big[\big\{F_n(s_t)-F(s_t)\big\}d C_t(s_t)K\big(\Lambda^{-\frac{1}{2}}(s-s_t)\big)\Big]\Big|\\
    &\le \mathcal{\xi}_1 |\Lambda^{-\frac{1}{2}}|\sup_{s_t}|F_n(s_t)-F(s_t)|,
\end{align*}

where $\xi$ is a constant and the last inequality can be derived by the assumption for the boundedness of $C_t(s_t)$ and $K(\cdot)$. By lemma 2.1 of \citet{schuster1969estimation}, we obtain that:$P_{S_t}\{\sup_{s_t}|F_n(s_t)-F(s_t)|>\epsilon\}\le \xi_2\exp(-2n\epsilon^2)$. Then:
\begin{align*}
    &P(\sup_s|\hat C_{t,1}(s)-E\{\hat C_{t,1}(s)\}|>\epsilon)\\
    &\le P(\xi_1|\Lambda^{-\frac{1}{2}}|\sup_{s_t}|F_n(s_t)-F(s_t)|>\epsilon)\\
    &=P(\sup_{s_t}|F_n(s_t)-F(s_t)|>\frac{\epsilon|\Lambda^{\frac{1}{2}}|}{\xi_1})\\
    &\le \xi_2\exp(-\frac{2n\epsilon^2|\Lambda|}{\xi_1^2}).
\end{align*}
Thus, if $2n|\Lambda|\to \infty$ as $n\to\infty$, then the first part of equation (\ref{twoparts}) converges to $0$. Equation (\ref{C1kernel}) is then proved. With similar proof for equation (\ref{pskernel}), we can obtain formula (\ref{Ckernel}). This ends the proof for formula (\ref{eq1}).


\subsubsection{Proof of Equation (\ref{eq2})}
First we write the left side of the equation as:
\begin{align*}
    &\sqrt{n}\mathbb P_n\Bigg[\Big\{L_2(H;\hat B,\hat C,\hat D)-L_1(H;\hat B,\hat C,\hat D)\phi_k^*\Big\}-\Big\{L_2(H;B,C,D)-L_1(H;B,C,D)\phi_k^*\Big\} \Bigg]\\
    &=\sqrt{n}\mathbb P_n \Big[\sum_{t=1}^{T-k+1} \{\hat M_{t,1}\hat M_{t,2}-M_{t,1}M_{t,2}\}\Big]\\
    &=\sum_{t=1}^{T-k+1}\sqrt{n}\mathbb P_n \Big[\{M_{t,1}(\hat M_{t,2}-M_{t,2})+M_{t,2}(\hat M_{t,1}-M_{t,1})+(\hat M_{t,1}-M_{t,1})(\hat M_{t,2}-M_{t,2})\}\Big],
\end{align*}

where,
\begin{gather*}
    \hat M_{t,1}=\begin{pmatrix}
    A_t^2-\hat B_t(S_{t})\\
    \{A_t-\hat C_t(S_{t})\}g_k(S_{t})
    \end{pmatrix},\\
    \hat M_{t,2}=Y_{t+k}-\hat D_t(S_{t}) - 
    \begin{pmatrix}
    A_t^2- \hat B_t(S_{t})\\
    \{A_t- \hat C_t(S_{t})\}g_k(S_{t})
    \end{pmatrix}^T\phi_k^*,\\
    M_{t,1}=\begin{pmatrix}
    A_t^2-  B_t(S_{t})\\
    \{A_t-  C_t(S_{t})\}g_k(S_{t})
    \end{pmatrix},\\
    M_{t,2}=Y_{t+k}- D_t(S_{t})- 
    \begin{pmatrix}
    A_t^2- B_t(S_{t})\\
    \{A_t- C_t(S_{t})\}g_k(S_{t})
    \end{pmatrix}^T\phi_k^*.
\end{gather*}

Thus, it is sufficient to show that:
\begin{gather}
    \sqrt{n}\mathbb P_n M_{t,1}(\hat M_{t,2}-M_{t,2})=o_p(1)\label{m1-m2hat},\\
    \sqrt{n}\mathbb P_n M_{t,2}(\hat M_{t,1}-M_{t,1})=o_p(1)\label{m1hat-m2},\\
    \sqrt{n}\mathbb P_n (\hat M_{t,1}-M_{t,1})(\hat M_{t,2}-M_{t,2})=o_p(1).\label{m1hat-m2hat}
\end{gather}
We first prove equation (\ref{m1-m2hat}). Let $G_{t,1}=A_t^2-B_t(S_t)$, $G_{t,2}=\{A_t-C_t(S_t)\}f_k(S_t)$, $G_{t,3}=Y_{t+k}-D_t(S_t)$ and $\hat G_{t,1}=A_t^2-\hat B_t(S_t)$, $\hat G_{t,2}=\{A_t-\hat C_t(S_t)\}f_k(S_t)$, $\hat G_{t,3}=Y_{t+k}-\hat D_t(S_t)$. Then equation (\ref{m1-m2hat}) can be written as:
\begin{gather*}
    \sqrt{n}\mathbb P_n \begin{pmatrix}
    G_{t,1}\\G_{t,2}
    \end{pmatrix}
    \Big\{\hat G_{t,3}-G_{t,3}+
    \begin{pmatrix}
    \hat G_{t,1}-G_{t,1}\\
    \hat G_{t,2}-G_{t,2}
    \end{pmatrix}^T\phi^*_k\Big\}
    =o_p(1).
\end{gather*}

Therefore, it is equivalent to show all the following equations :
\begin{gather}
    \sqrt{n}\mathbb P_n G_{t,1}\{\hat G_{t,3}-G_{t,3}\}=o_p(1);\nonumber\\
    \sqrt{n}\mathbb P_n G_{t,2}\{\hat G_{t,3}-G_{t,3}\}=o_p(1);\nonumber\\
    \sqrt{n}\mathbb P_n G_{t,1}\{\hat G_{t,1}-G_{t,1}\}=o_p(1);\nonumber\\
    \sqrt{n}\mathbb P_n G_{t,2}\{\hat G_{t,2}-G_{t,2}\}=o_p(1);\nonumber\\
    \sqrt{n}\mathbb P_n G_{t,1}\{\hat G_{t,2}-G_{t,2}\}=o_p(1);\nonumber\\
    \sqrt{n}\mathbb P_n G_{t,2}\{\hat G_{t,1}-G_{t,1}\}=o_p(1).\label{g-equations}
\end{gather}
We show the proof of the last equation above. The rest of the equations can be proved similarly. First write it as:
\begin{align*}
    &\sqrt{n}\mathbb P_n G_{t,2}\{\hat G_{t,1}-G_{t,1}\}\\
    &=-\sqrt{n}\mathbb P_n \{A_t-C_t(S_t)\}\{\hat B_t(S_t)-B_t(S_t)\}.
\end{align*}
%
Let $\hat B_{t,1}(s)=\sum_{j=1}^n {A_t^j}^2K_\Lambda(S_t^j-s)/n$, $B_{t,1}(s)=B_{t}(s)f_{S_t}(s)$. Then $\hat B_t(s)=\hat B_{t,1}(s)/\hat f_{S_t}(s)$. 
If we can obtain that
\begin{align}
    \lim_{n\to \infty}\text{Var}\Big\{\sqrt{n|\Lambda^{1/2}|}\Big(\hat B_{t,1}(s)-B_{t}(s)\hat f_{S_t}(s)\Big)\Big\}<\infty,
\end{align} then from appendix B.1 of \citet{zhu2019}, we obtain that:  under the assumptions:
$\sqrt{n|\Lambda^{1/2}|}\Big(\hat B_{t,1}(s)-B_{t}(s)\hat f_{S_t}(s)\Big)$ converge in distribution to a mean $0$ normal distribution. Together with equation \ref{pskernel} and 
the assumption that $f_{S_t}(s)$ is bounded away from $0$, we can obtain that,
\begin{align*}
    &\sqrt{n|\Lambda|^{\frac{1}{2}}}\{\hat B_t(S_t)-B_t(S_t)\}\\&
    =\sqrt{n|\Lambda|^{\frac{1}{2}}}\Big\{\frac{\hat B_{t,1}(S_t)-B_t(S_t)\hat f_{S_t}(S_t)}{f_{S_t}(S_t)}\Big\}+o_p(1)\\&
    =\sqrt{n|\Lambda|^{\frac{1}{2}}}\Big\{\frac{1}{n}\sum_{j=1}^n\tilde B_t^j(S_t)K_\Lambda(S_t^j-S_t)\Big\}+o_p(1),
\end{align*} 
where $\tilde B_t^j(s)=\{A_t^2-E(A_t^2|S_t=s)\}/f_{S_t}(s)$.

Then:
\begin{align*}
    &\sqrt{n}\mathbb P_n G_{t,2}\{\hat G_{t,1}-G_{t,1}\}\\
    &=-\sqrt{n}\mathbb P_n \{A_t-C_t(S_t)\}\{\hat B_t(S_t)-B_t(S_t)\}\\
    &=-\frac{1}{n\sqrt{|\Lambda^{\frac{1}{2}}|}}\sum_{i=1}^n\Big\{A_t^i-C_t(S_t^i)\Big\}
        \Big[\sqrt{n|\Lambda^{\frac{1}{2}}|}\Big\{\hat B_t(S_t^i)-B_t(S_t^i)\}\Big]\\
    &=-\frac{1}{n\sqrt{|\Lambda^{\frac{1}{2}}|}}\sum_{i=1}^n\Big\{A_t^i-C_t(S_t^i)\Big\}
    \Big[\sqrt{n|\Lambda^{\frac{1}{2}}|}\Big\{\frac{1}{n}\sum_{j=1}^n\tilde B_t^j(S_t^i)K_\Lambda(S_t^j-S_t^i)\Big\}+o_p(1)\Big]\\
    &=-\frac{\sqrt{n}}{n^2}\sum_{i=1}^n\Big\{A_t^i-C_t(S_t^i)\Big\}\Big\{\sum_{j=1}^n\tilde B_t^j(S_t^i)K_\Lambda(S_t^j-S_t^i)\Big\}+o_p(1)\\
    &=-\frac{\sqrt{n}}{n^2}\sum_{i=1}^n\sum_{j=1}^n\Big\{A_t^i-C_t(S_t^i)\Big\}\tilde B_t^j(S_t^i)K_\Lambda(S_t^j-S_t^i)+o_p(1).
\end{align*}

The third equation above is based on $\sqrt{n|\Lambda^{\frac{1}{2}}|}\to \infty$ and $\frac{\sqrt{n}}{n}\sum_{i=1}^n(A_t^i-C_t(S_t^i))\xrightarrow{d}N(0, E\{\text{Var}(A_t|S_t\}).$
Thus we just need to prove that the first term is $o_p(1)$. 
The first term above is a $\sqrt{n}$ times a U-statistic plus an $o_p(1)$ term when written as:
\begin{align*}
    &\frac{\sqrt{n}}{n^2}\sum_{j=1}^n \sum_{i<j}\Bigg[ \tilde B_t^j(S_{t}^i)  K_{\Lambda}(S_{t}^j-S_{t}^i) \{A^i_{t}-C_t(S^i_{t})\}
    +\tilde B_t^i(S_{t}^j)  K_{\lambda}(S_{t}^j-S_{t}^i) \Big\{A_t^j-C_t(S^j_{t})\Big\}\Bigg]\\
    &+\frac{1}{n}\Bigg[\frac{\sqrt{n}}{n}\sum_{i=1}^n \tilde B_t^i(S_{t}^i)\Big\{A^i_{t}-C_t(S^i_{t})\Big\}\Bigg].
\end{align*}

The second term above is $o_p(1)$ because of the law of large numbers. 
The expectation of the U-statistics is equal to :
\begin{align*}
    &\frac{n-1}{n}E\Big[\tilde B_t^j(S_{t}^i)  K_{\Lambda}(S_{t}^j-S_{t}^i) \Big\{A^i_{t}-C_t(S^i_{t})\Big\}\Big]\\
    &=\frac{n-1}{n}E\Big[\frac{{A_t^j}^2-E(A_t^2|S_t=S_t^i)}{f_{S_t}(S_t^i)}  K_{\Lambda}(S_{t}^j-S_{t}^i) \Big\{A^i_{t}-E(A_{t}|S_t=S^i_{t})\Big\}\Big]\\
    &=\frac{n-1}{n}E\Bigg(E\Big[ \Big\{A^i_{t}-E(A_{t}|S_t=S^i_{t})\Big\}\Big|S_t^i, A_t^j, S_t^j \Big]\frac{{A_t^j}^2-E(A_t^2|S_t=S_t^i)}{f_{S_t}(S_t^i)}  K_{\lambda}(S_{t}^j-S_{t}^i)\Bigg)\\
    &=0.
\end{align*}

By the properties of U-statistics, the variance of $\sqrt{n}$ times the U-statistics converge to:
\begin{align*}
    & \text{Var}\Big\{E\Big[ \tilde B_t^j(S_t^i) K_{\Lambda}(S^i_{t}-S^j_{t})\{A^i_{t}
    -C_t(S^i_{t})\}+
    \tilde B_t^i(S^j_{t}) K_{\Lambda}(S^j_{t}-S^i_{t})\{A^j_{t}-
    C_t(S^i_{t})\}\Big|S^i_{t}, A^i_{t}\Big] \Big\}.
\end{align*}

We can obtain:
\begin{align*}
    &E\Big[ \tilde B_t^j(S^i_{t}) K_{\Lambda}(S^j_{t}-S^i_{t})\{A^i_{t}-C_t(S^i_{t})\}+\tilde B_t^i(S^j_{t}) K_{\lambda}(S^i_{t}-S^j_{t})\{A^j_t-C_t(S^j_{t})\}\Big|S^i_{t}, A^i_{t}\Big]\\
    &=E\Big[\tilde B_t^j(S^i_{t}) K_{\Lambda}(S^j_{t}-S^i_{t})\{A^i_{t}-C_t(S^i_{t})\}|S^i_{t}, A^i_{t}\Big]\\
    &=E\Big[\{{A_t^j}^2-B_t(S_t^i)\} K_{\Lambda}(S^j_{t}-S^i_{t})|S^i_{t}, A^i_{t}\Big]\frac{A^i_{t}-C_t(S^i_{t})}{f_{S_t}(S_t^i)}.
\end{align*}

From calculation in section \ref{section:basic_calc}, we can obtain that:
\begin{align*}
    \sup_s|E\{{A_t^j}^2 K_\Lambda(S_t^j-s)\}-B_t(s)f_{S_t}(s)|=O(|\Lambda^{\frac{1}{2}}|) \\
    \sup_s|E\{ K_\Lambda(S_t^j-s)\}-f_{S_t}(s)|=O(|\Lambda^{\frac{1}{2}}|) .
\end{align*}

Thus,
\begin{align*}
    &E\Big[\{{A_t^j}^2-B_t(S_t^i)\} K_{\Lambda}(S^j_{t}-S^i_{t})|S^i_{t}, A^i_{t}\Big]\le \xi_3|\Lambda|^{\frac{1}{2}}.
\end{align*}
for some constant $\xi_3$. Thus,
\begin{align*}
    &\text{Var}\Big\{E\Big[ \tilde B_t^j(S_t^i) K_{\Lambda}(S^i_{t}-S^j_{t})\{A^i_{t}
    -C_t(S^i_{t})\}+
    \tilde B_t^i(S^j_{t}) K_{\Lambda}(S^j_{t}-S^i_{t})\{A^j_{t}-
    C_t(S^i_{t})\}\Big|S^i_{t}, A^i_{t}\Big] \Big\}\\
    &\le \xi^2|\Lambda|\text{Var}\{\frac{A_t^i-C_t(S_t^i)}{f_{S_t}(S_t^i)}\}.
\end{align*}
Then as long as $\text{Var}\{(A_t^i-C_t(S_t^i))/f_{S_t}(S_t^i)\}<\infty$, the variance of the U-statistics converges to 0. Since we assumed that $f_{S_t}(s)$ is bounded away from $0$  and $E(A_t^2|S_t=S_t^i)<\infty$, this conditional can be satisfied. 
Thus, both the expectation and the variance of the $\sqrt{n}$ times the U-statistics converge to $0$,  so $\sqrt{n}$ times the U-statistic converges in probability to $0$.  Thus equation (\ref{g-equations}) can be proved. With similar proof for the other equations above equation (\ref{g-equations}), we can obtain equation (\ref{m1-m2hat}). Equation (\ref{m1hat-m2}) can be proved similarly. Equation (\ref{m1hat-m2hat}) can be proved with similar calculation. We omit the details here due to the length of the proof. This completes the proof for equation (\ref{eq2}).

\subsection{Proof for Equation (\ref{eq3})}
Equation \ref{eq3} can be simply obtained with the central limit theorem. Thus the proof for theorem is completed.

\subsection{Form for lag k effect under the simulation setting}
The true value for the lag $2$ effect is:
\begin{align*}
  &E(Y_{t+2}|A_{t}=a, S_{t})-E(Y_{t+2}|A_{t}=0, S_{t})=\\
& -(\tau_1\eta_2+\tau_2-\beta_1 \eta_2)(\tau_2+\tau_1\eta_2) a^2 +
\Big\{ \theta_1\eta_2 +\theta_2 +\beta_0(\tau_1\eta_2+\tau_2)\Big\}a\\
&+\Big\{(\tau_1\eta_2+\tau_2)(-2\tau_1\eta_1+\beta_1\eta_1)+\beta_1\tau_1\eta_1\eta_2\Big\}aX_t.
\end{align*}

For $k\ge 3$:
If we have:
\begin{align*}
    &E(Y_{t+k-1}|A_{t}=a,S_{t})=\alpha_{k-1,1} X_{t}+\alpha_{k-1,2} X_{t}^2 +
    \alpha_{k-1,3} A_{t}^2 +\alpha_{k-1,4} A_{t}+\alpha_{k-1,5} A_{t}X_{t}.
\end{align*}

Then
\begin{align*}
    &E(Y_{t+k}|A_{t}=a,S_{t})\\
    &=\alpha_{k-1,1} X_{t+1}+\alpha_{k-1,2} X_{t+1}^2 +
    \alpha_{k-1,3} A_{t+1}^2 +\alpha_{k-1,4} A_{t+1}+\alpha_{k-1,5} A_{t+1}X_{t+1}\\
    &=\alpha_{k-1,1} (\eta_1 X_t +\eta_2 A_t)+\alpha_{k-1,2} (\eta_1 X_t +\eta_2 A_t)^2 +
    \alpha_{k-1,3} \Big\{\tau_1\eta_1 X_t +(\tau_1\eta_2+\tau_2)A_t\Big\}^2 \\
    &+\alpha_{k-1,4} \Big\{\tau_1\eta_1 X_t +(\tau_1\eta_2+\tau_2)A_t\Big\}+\alpha_{k-1,5} \Big\{\tau_1\eta_1 X_t
    +(\tau_1\eta_2+\tau_2)A_t\Big\}(\eta_1 X_t +\eta_2 A_t)\\
    &=\alpha_{k,1} X_{t}+\alpha_{k,2} X_{t}^2 +\alpha_{k,3} A_{t}^2 +\alpha_{k,4} A_{t}+\alpha_{k,5} A_{t}X_{t},
\end{align*}

where 
\begin{align*}
    &\alpha_{k,1}=\Big\{\alpha_{k-1,1} +\alpha_{k-1,4}\tau_1\Big\}\eta_1,\\
    &\alpha_{k,2}=\Big\{\alpha_{k-1,2}+\alpha_{k-1,3} \tau_1^2 +\alpha_{k-1,5} \tau_1\Big\}\eta_1^2,\\
    &\alpha_{k,3}=\Big\{\alpha_{k-1,2}\eta_2^2+\alpha_{k-1,3}(\tau_1\eta_2+\tau_2)^2+\alpha_{k-1,5}(\tau_1\eta_2
    +\tau_2)\eta_2\Big\},\\
    &\alpha_{k,4}=\Big\{\alpha_{k-1,1}\eta_2+\alpha_{k-1,4}(\tau_1\eta_2+\tau_2)\Big\},\\
    &\alpha_{k,5}=\Big\{2\eta_1\eta_2\alpha_{k-1,2}+\alpha_{k-1,3} 2\tau_1\eta_1(\tau_1\eta_2+\tau_2)+\\&
    \alpha_{k-1,5}\big[\tau_1\eta_1\eta_2+\eta_1(\tau_1\eta_2+\tau_2)\big]\Big\}.
\end{align*}

Then lag k effect is:
\begin{align*}
    \alpha_{k,3}A_t^2+\alpha_{k,4}A_t+\alpha_{k,5}A_tX_t.
\end{align*}

\subsection{Proof of Assumption (\ref{assumption:strong}) under the Simulation Setting}
\label{section:mhealth-simulation-assumption-proof}
According to the data generation model for our simulation setting,  $$Y_{t+1}(\bar A_{t-1},a_t=a)=\theta_1 X_t+\theta_2 A_{t-1}-a(a-\beta_0-\beta_1X_t)+\epsilon_{t+1}.$$  
$$A_t\sim \text{Normal}(\tau_1X_t+\tau_2A_{t-1}).$$
When $\theta_2=0$, 
$$Y_{t+1}(\bar A_{t-1},a_t=a)=\theta_1 X_t-a(a-\beta_0-\beta_1X_t)+\epsilon_{t+1}.$$
is independent of $A_t$ given $S_t=X_t$. Thus assumption (\ref{assumption:strong}) is satisfied for $k=1$. However, when $\theta_2\neq 0$, this assumption is not satisfied for $k=1$.

For $k=2$, first since $X_{t+1}(\bar A_{t-1},a_t=2)\sim \text{Normal} (\eta_1 X_t+\eta_2 A_t,\sigma^2)$, it is trivial to see that:
\begin{equation}\label{equation:x_t+1}
    X_{t+1}(\bar A_{t-1},a_t=a)\perp A_t |X_t.
\end{equation} Since 
$$A_{t+1}(\bar A_{t-1},a_t=a)\sim \text{Normal}\Big(\tau_1 X_{t+1}(\bar A_{t-1},a_t=a)+\tau_2 a_t, \sigma^2\Big),$$ 
we can obtain that 
\begin{equation}\label{equation:a_t+1}
    A_{t+1}(\bar A_{t-1},a_t=a)\perp A_t |X_t.
\end{equation}
Therefore, 
\begin{align*}
    &Y_{t+2}(\bar A_{t-1}, a_t=a, A_{t+1}^{a_t=a})=\\
    &\theta_1 X_{t+1}(\bar A_{t-1},a_t=a)+\theta a -A_{t+1}(\bar A_{t-1},a_t=a) \Big\{A_{t+1}(\bar A_{t-1},a_t=a)-\beta_0-\\&
    \beta_1X_{t+1}(\bar A_{t-1},a_t=a)\Big\}+\epsilon_{t+2}
\end{align*} is independent of $A_t$ given $X_t$ based on Equation (\ref{equation:x_t+1}) and (\ref{equation:a_t+1}). This is true even when $\theta_2\neq 0$. 

Using induction, we can also prove that for any $k\ge 3$, 
$$Y_{t+k}(\bar A_{t-1},a_t=a,A_{t+1},\dots,A_{t+k-1})\perp A_t|X_t.$$

\subsection{Additional Simulation Results}
\label{section:mhealth:appendix-k23}
The true parameters for $k=2,3$ when $\theta_2=-0.1$ are :  $(\alpha_2,\beta_{2,0},\beta_{2,1})=(-0.21,0.06,-0.08)$; $(\alpha_3,\beta_{3,0},\beta_{3,1})= (-0.0125,-0.05,-0.03)$. Table \ref{table:mhealth-appendix-k23} presents the result for the estimated parameters when $S_t=X_t$. As shown in the table, the estimated parameters appeared to be unbiased.
\begin{table}
\small \renewcommand{\arraystretch}{1}
\begin{center}
 \caption{Simulation results from $200$ replicates when $\theta_2=-0.1$.\label{table:mhealth-appendix-k23}}
\begin{threeparttable}
 \begin{tabular}{c| c| c c c c| c c c c| c c c c}
\toprule
    \multicolumn{2}{c|}{}& \multicolumn{4}{c|}{$\alpha_k$}& \multicolumn{4}{c|}{$\beta_{k,0}$}& \multicolumn{4}{c}{$\beta_{k,1}$}\\
\midrule
$k$    & n   & Bias$^1$ & SD$^1$ & SE$^1$ & CP & Bias$^1$  & SD$^1$    & SE$^1$ & CP & Bias$^1$  & SD$^1$    & SE$^1$ & CP\\
\midrule
 2       & 100 & 1.9 &31.4 & 29.2 & 91.5     & -1.2 & 23.4 & 22.4 & 93.5   & -4.7  & 79.1 & 68.3  & 93.0 \\
         & 200 & -1.0 &23.8 & 20.9 & 91.5     & -0.1 & 16.6 & 15.8 & 93.0   & 3.6  & 56.3 & 47.9  & 90.5 \\
         & 400 & -1.1&14.8 & 14.8 & 95.5     & 1.0 & 11.9  & 11.1 & 93.5   & 1.0  & 33.6 & 33.2  & 95.0 \\
 \midrule
 3       & 100 & 2.0  &32.2  & 26.8 & 88.5     & 4.2  & 22.2 &  21.1 & 94.0 &  2.9 & 75.1 & 67.1  & 90.5    \\
         & 200 & -3.1 &19.5  & 18.8 & 93.0     & 0.6  & 15.6 &  14.7 & 91.5&   5.3 & 50.8 & 45.7  & 92.0   \\
         & 400 & 1.1 &15.7  & 13.3 & 89.5     &  0.7  & 10.8 &  10.3 & 94.0&  -2.5 & 36.7 & 31.5  & 92.0   \\
\bottomrule
\end{tabular}
\begin{tablenotes}
\footnotesize{
\item[1] Note: These columns are in $10^{-3}$ scale
\item[2] Note: SD refers to the standard deviation of the estimated parameters from $200$ replicates, SE refers to the mean of the estimated standard errors calculated by our covariance function, CP refers to the coverage probability of the $95\%$ confidence intervals calculated using the estimated standard errors.}
\item[3] Note: The worst case Monte Carlo standard error for proportions is $2.3\%$.
\end{tablenotes}
\end{threeparttable}
\end{center}
\end{table}

For $k=1$, we can correct the bias by estimating the parameters with $S_t=(X_t,A_{t-1})$. The model for the lag $1$ treatment effect is thus: $\tau_{t,1}=\alpha_1 a^2+(\beta_{1,0}a+\beta_{1,1}X_t+\beta_{1,2}A_{t-1})a$. The true parameters are: $(\alpha_1,\beta_{1,0},\beta_{1,1},\beta_{1,2})=(-1,0,2,0)$. Since the dimension of $S_t$ has increased, we use the bandwidth $\lambda_j=n^{-1/4}\text{sd}(S_{t,j})$. The estimated parameters are presented in Table \ref{table:mhealth-appendix-sim-corrected}. From the results we see that the estimated parameters appeared to be unbiased. However, the estimated standard deviation was smaller than the actual standard deviation, leading to lower coverage probability when sample size was small. This implies that when the dimension of covariates increases, the estimated standard error converges slower to the actual standard deviation.

\begin{table}
\small \renewcommand{\arraystretch}{1}
\begin{center}
 \caption{Simulation results from $200$ replicates when $\theta_2=-0.1$.\label{table:mhealth-appendix-sim-corrected}}
\begin{threeparttable}
 \begin{tabular}{c| c|c  c c c c}
\toprule
$k$  & n & Parameter     & Bias$^1$ & SD$^1$ & SE$^1$ & CP \\
\midrule
 1  &100 &$\alpha_k$   & 18.3 & 23.4 & 21.8& 83.5 \\
    &    &$\beta_{k,0}$& 11.7 & 15.9 & 14.8  & 82.0\\
    &    &$\beta_{k,1}$& -20.5& 59.3 & 54.5 & 91.5\\
    &    &$\beta_{k,2}$& 12.8 & 32.9 & 31.2 & 90.5\\\cline{2-7}
    &200 &$\alpha_k$   &9.2   & 13.8 & 15.0& 92.5 \\
    &    &$\beta_{k,0}$&6.6 & 12.2 & 10.3 & 85.5\\
    &    &$\beta_{k,1}$&-5.2 & 35.5 & 37.5 & 97.0\\
    &    &$\beta_{k,2}$&4.6 & 21.1 & 21.3 & 95.5\\\cline{2-7}
    &400 &$\alpha_k$   &5.3 & 11.0  & 10.5 & 92.0 \\
    &    &$\beta_{k,0}$&3.4 & 8.1 & 7.3 & 88.0\\
    &    &$\beta_{k,1}$&-2.6 & 28.4 & 26.1 & 93.0\\
    &    &$\beta_{k,2}$&1.7 & 14.5 & 14.9 & 96.0\\
\bottomrule
\end{tabular}
\begin{tablenotes}
\footnotesize{
\item[1] Note: These columns are in $10^{-3}$ scale
\item[2] Note: SD refers to the standard deviation of the estimated parameters from $200$ replicates, SE refers to the mean of the estimated standard errors calculated by our covariance function, CP refers to the coverage probability of the $95\%$ confidence intervals calculated using the estimated standard errors.}
\item[3] Note: The worst case Monte Carlo standard error for proportions is $2.3\%$.
\end{tablenotes}
\end{threeparttable}
\end{center}
\end{table}

\subsection{Additional Results for Ohio Type 1 Diabetes Dataset}
We applied the proposed method for the Ohio Type 1 Diabetes Dataset. The dataset consists of data from 6 patients. The result for patient 6 has been presented in the article. In Table \ref{table:mhealth-realdata-additional}, we present the additional results from the other patients. It is likely that the decision process for insulin dosage is different for each patient. Thus using the same set of $S_t$ for all patients might not be the optimal choice.

\begin{table}
\centering\footnotesize
 \caption{Estimated variables with the Ohio type 1 diabetes dataset\label{table:mhealth-realdata-additional}}
\begin{threeparttable}
 \begin{tabular}{p{1cm}|p{2cm} p{2cm} p{2cm} p{2cm} p{2cm}}
\toprule
    \multicolumn{6}{c}{Patient 1}\\
    \hline
    $k$ &1&2&3&4&Weighted \\
    \hline
$\alpha_k$&17(7.6)&21.8(10.6)&16.6(10.6)&10.9(12.7)&16(9.1)\\
$\beta_{k,0}$&-172.3(73.1)&-92.3(103.9)&-11(112.4)&-50(116.4)&-63.2(86.3)\\
$\beta_{k,1}$&1.4(0.7)&3.1(1.4)&3.6(1.6)&3.3(1.7)&3.1(1.3)\\
$\beta_{k,2}$&-0.8(3.1)&2.1(4.7)&5.9(4.8)&3.9(5.9)&2.9(3.6)\\
$\beta_{k,3}$&-9.1(51.5)&-104.4(80.4)&-93.7(88.3)&34.9(66.7)&-59.6(56.9)\\
$\beta_{k,4}$&12.9(8.9)&2.6(19.2)&-11.6(23)&-170(35.2)&-110.4(33.6)\\
\bottomrule
\end{tabular}

 \begin{tabular}{p{1cm}|p{2cm} p{2cm} p{2cm} p{2cm} p{2cm}}
\toprule
    \multicolumn{6}{c}{Patient 2}\\
    \hline
    $k$ &1&2&3&4&Weighted \\
    \hline
$\alpha_k$&0.2(0.7)&0.5(0.8)&0.6(1)&-0.8(1.1)&0.1(0.8)\\
$\beta_{k,0}$&-8.8(11.1)&14.1(18)&18(22.6)&14.7(24.2)&10.7(17.8)\\
$\beta_{k,1}$&0.6(0.3)&-0.2(0.2)&0(0.2)&-0.1(0.5)&0.1(0.2)\\
$\beta_{k,2}$&0.4(0.1)&0.2(0.3)&0.1(0.3)&0.3(0.3)&0.3(0.2)\\
$\beta_{k,3}$&-2.6(6.8)&-19.1(16.1)&-22.2(22.5)&-5.3(23.8)&-13.1(16.9)\\
$\beta_{k,4}$&-2.1(2.2)&-3(4)&-4.6(4.5)&-6.3(5.2)&-4.2(4)\\
\bottomrule
\end{tabular}
 \begin{tabular}{p{1cm}|p{2cm} p{2cm} p{2cm} p{2cm} p{2cm}}
\toprule
    \multicolumn{6}{c}{Patient 3}\\
    \hline
    $k$ &1&2&3&4&Weighted \\
    \hline
$\alpha_k$&0.6(1)&-2.2(1.8)&-5(2.6)&-4.6(2.9)&-2.9(1.9)\\
$\beta_{k,0}$&-14(39.7)&69.2(57.3)&159.8(81.9)&173.5(88.3)&98.8(63.9)\\
$\beta_{k,1}$&0.1(0.1)&0.4(0.3)&0.6(0.4)&0.7(0.4)&0.4(0.3)\\
$\beta_{k,2}$&0.1(0.1)&0.2(0.2)&0.4(0.3)&0.3(0.3)&0.3(0.2)\\
$\beta_{k,3}$&-11.2(46.6)&-78.1(67.3)&-148.4(93.1)&-166(98.1)&-103.3(74.8)\\
$\beta_{k,4}$&2.3(2.5)&3.4(2.9)&5.9(3.3)&7.5(3.2)&4.8(2.7)\\
\bottomrule
\end{tabular}
 \begin{tabular}{p{1cm}|p{2cm} p{2cm} p{2cm} p{2cm} p{2cm}}
\toprule
    \multicolumn{6}{c}{Patient 4}\\
    \hline
    $k$ &1&2&3&4&Weighted \\
    \hline
$\alpha_k$&-1.2(5.1)&-7.4(5.9)&-6.5(4.8)&-5.8(4.6)&-6.2(4.4)\\
$\beta_{k,0}$&-129(138.1)&162.5(95.5)&154.2(124.4)&167.4(104.2)&97.8(72.4)\\
$\beta_{k,1}$&0.2(0.4)&0.1(0.5)&-0.8(0.7)&-0.4(1)&-0.1(0.5)\\
$\beta_{k,2}$&0.4(0.5)&1.3(0.9)&1.2(0.8)&1.2(0.8)&0.7(0.6)\\
$\beta_{k,3}$&130.6(169.5)&-173.4(130.2)&-146.4(162.6)&-182.3(143.9)&-96(97)\\
$\beta_{k,4}$&1.2(3.5)&-8.2(10.1)&-13.9(12.3)&-21.2(9.7)&-10.3(7.7)\\
\bottomrule
\end{tabular}
 \begin{tabular}{p{1cm}|p{2cm} p{2cm} p{2cm} p{2cm} p{2cm}}
 \toprule
    \multicolumn{6}{c}{Patient 5}\\
    \hline
    $k$ &1&2&3&4&Weighted \\
    \hline
$\alpha_k$&2.7(2.9)&6.7(4.4)&6.6(4.9)&2.7(5.2)&3.9(3.7)\\
$\beta_{k,0}$&-271(110.8)&-514.8(139.1)&-321(160.8)&-110.6(151.9)&-303.4(127.8)\\
$\beta_{k,1}$&0.4(0.3)&1.2(0.7)&0.1(0.7)&-0.6(0.8)&0.3(0.6)\\
$\beta_{k,2}$&0.6(0.3)&0.5(0.6)&0.5(0.7)&0.8(0.7)&0.6(0.5)\\
$\beta_{k,3}$&173.1(87.1)&324.2(119.8)&201.9(134.1)&78.7(125.6)&198.1(103.7)\\
$\beta_{k,4}$&28.8(13)&12.4(10.8)&-5.1(6)&-1.6(13.2)&8.4(7.8)\\
     \bottomrule
\end{tabular}
\begin{tablenotes}
\footnotesize{
\item[1] Note: These columns are in $10^{-2}$ scale .
\item[2] Note: The numbers in the parenthesis are the estimated standard errors calculated by the covariance formula.
\item[3] Note: The last column presents the estimated parameters for the lag $4$ weighted advantage with $w_1=w_2=w_3=w_4=1/4$.
}
\end{tablenotes}
\end{threeparttable}
\end{table}

\end{document}